\documentclass[a4paper,fleqn,usenatbib]{mnras}

\usepackage[T1]{fontenc}
\usepackage{ae,aecompl}
\usepackage{graphicx}	
\usepackage{amsmath}	
\usepackage{amssymb}	
\usepackage{tablefootnote}
\usepackage{lscape}

\title[]{A Hubble Space Telescope Survey of the Host Galaxies of Superluminous Supernovae}

\author[C. R. Angus et al. ]{C. R. Angus$^{1}$,\thanks{E-mail: C.R.Angus@warwick.ac.uk} A. J. Levan$^{1}$, D. A. Perley$^{2,3}$, N. R. Tanvir$^{4}$,  J.D. Lyman$^{1}$, 
\newauthor  E.R. Stanway$^{1}$, A.S. Fruchter$^{5}$\\
$^{1}$Department of Physics, University of Warwick, Coventry CV4 7AL, UK.\\
$^{2}$Dark Cosmology Centre, Niels Bohr Institute, University of Copenhagen, Juliane Maries Vej 30, DK-2100 Copenhagen, Denmark.\\
$^{3}$Department of Astronomy, California Institute of Technology, Pasadena, CA, 91106, USA.\\
$^{4}$Department of Physics and Astronomy, University of Leicester, Leicester LE1 7RH, UK.\\ 
$^{5}$Space Telescope Science Institute, 3700 San Martin Drive, Baltimore, MD 21218, USA.}

\pubyear{2016}

\begin{document}
\label{firstpage}
\pagerange{\pageref{firstpage}--\pageref{lastpage}}
\maketitle

\begin{abstract}
We present Hubble Space Telescope (HST) WFC3 UV and near-IR (nIR)
imaging of 21 Superluminous Supernovae (SLSNe) host galaxies, providing
a sensitive probe of star formation and stellar mass with the hosts.  
Comparing the photometric and
morphological properties of these host galaxies with those of core
collapse supernovae (CCSNe) and long-duration gamma-ray bursts
(LGRBs), we find SLSN hosts are fainter and more compact at both
UV and nIR wavelengths, in some cases we barely recover hosts
with absolute magnitude around MV $\approx$ -14. With the addition
of ground based optical observations and archival results, we produce
spectral energy distribution (SED) fits to these hosts, and show
that SLSN hosts possess lower stellar mass and star formation rates.
This is most pronounced for the hydrogen deficient Type-I SLSN
hosts, although Type-II H-rich SLSN host galaxies remain distinct
from the bulk of CCSNe, spanning a remarkably broad range of absolute
magnitudes, with $\sim$30$\%$ of SLSNe-II arising from galaxies fainter
than M$_{nIR}$ $\sim$ -14. The detection of our faintest SLSN hosts increases the
confidence that SLSNe-I hosts are distinct from those of LGRBs
in star formation rate and stellar mass, and suggests that apparent
similarities in metallicity may be due to the limited fraction
of hosts for which emission line metallicity measurements are
feasible. The broad range of luminosities of SLSN-II hosts is
difficult to describe by metallicity cuts, and does not match the
expectations of any reasonable UV-weighted luminosity function,
suggesting additional environmental constraints are likely necessary
to yield hydrogen rich SLSNe.
\end{abstract}


\section{Introduction}
During the past decade, time resolved, wide field, transient surveys such as the Panoramic Survey Telescope And Rapid Response System (Pan-STARRS, \citealt{Pan-STARRS_REF}), Palmomar Transient Factory (PTF, \citealt{PTF_REF}) and Catalina Real Time Survey (CRTS, \citealt{Drake2009}), have revealed the extent of diversity amongst cosmic explosions showing that the optical transient sky exhibits a much broader range of 
of events in both luminosity and duration than spanned by classical supernovae. These discoveries have largely been possible thanks to the unprecedented combination of depth, areal coverage and cadence of observations that are provided by such surveys, enabling order of magnitude increases in the number of transients recorded. {\color{black} This is combined with increasingly effective and sophisticated follow-up, that has allowed} rare, hitherto unrecognised, populations of events to be uncovered, {\color{black} and sufficient numbers of events to be located to identify new populations, rather than just extreme outliers}. Of particular interest
are populations of highly luminous, but extremely rare supernovae, peaking at magnitudes brighter than $M_V \sim -21$, a factor of $\sim 100$ times brighter than the majority of core collapse supernovae, and 10 times brighter at peak than SNe Ia. The achievement of such high luminosities during stellar collapse is likely a result of peculiar and poorly understood explosion mechanisms, through which we may shed light upon the exotic stars from which they originate. These Superluminous Supernovae (SLSNe) have been observed since  at least the mid-1990's \citep{Knop1999}, but it is only in the past few years that sufficient numbers with detailed follow-up have become available, enabling them to be identified as a new population of events \citep{Quimby2011} complete with internal diversity similar to that seen in normal SNe, in particular the hydrogen-rich and -poor dichotomy \citep[e.g.][]{Gal-Yam2012}. 
In addition to their high luminosity, SLSNe also  frequently exhibit long rise times, remaining brighter than the peak magnitude of most SNe for hundreds of days. Furthermore, they are extremely blue, in contrast to many SNe whose UV-light vanishes due to metal line blanketing shortward of $\sim$ 3000$\AA$ in the rest frame. The combination of luminosity, longevity and blue colours makes them potentially powerful cosmological probes, visible, even with current technology, out to $z>4$ \citep{Inserra2014}. Indeed, the most distant SNe detected\footnote{We exclude GRBs from this definition since while they are known to be core collapse events we do not directly detect the SNe light beyond $z \sim 1$} are of the SLSNe variety \citep{Howell2013,Cooke2012}

SLSNe have been broadly classified in a similar manner to normal CCSNe, with hydrogen-poor and hydrogen-rich events being labelled Type-I and Type-II, respectively. Independent of these spectroscopic classifications, there is a suggested population of hydrogen poor events whose light curves appear to be shaped by the rate of radioactive decay of nickel ($^{56}$Ni), Type-R \citep{Gal-Yam2012}.  At face value, if powered by standard radioactivity, the peak brightness of SLSN events imply the synthesis of several solar masses of  $^{56}$Ni during the explosion, a feat impossible in the most massive Galactic stars. This is due to ongoing mass-loss processes throughout their Main Sequence lifetime that inhibit the growth of the core so that 
it cannot reach masses greater than $\sim$60 M$_{\odot}$  \citep{Heger2002}, the mass needed to synthesise the levels of $^{56}$Ni implied by 
SLSNe peak magnitudes. This has led to suggestions that SLSNe originate from massive (M$\textgreater$100M$_{\odot}$), low metallicity stars,
 that may bear a strong resemblance to first generation, population III stars \citep[e.g.][]{Gal-Yam2009}.  In this case, the explosion
mechanism may be the complete destruction of the core in long sought after Pair Instability Supernovae (PISNe, e.g. \citealt{Rakavy1967}), whereby the production of electron-positron pairs within the core rapidly reduces internal pressure, causing the star to collapse. The resulting thermonuclear detonation disrupts the star entirely, leading to a luminous outburst. While mooted as a possible origin for many SLSNe, the Type-R events now seem the most 
likely candidates for such explosions.

However, this interpretation remains controversial with some arguing that they are simply a subset of Type-I SLSNe for which
the similarity with a nickel decay is coincidental \citep{Inserra2013A,Nicholl2013}. There are various proposed mechanisms for the production of such luminous outbursts. An extremely bright supernova may be produced via the interaction of the SN shock wave with a dense shell of material expelled from the star during a prior evolutionary phase, shock heating the hydrogen rich material, causing it to luminesce over a larger radius (the interaction model, \citealt{Chevalier2011}). Otherwise, the re-energizing of the SN shock wave via accretion onto a compact object or the spin down of a magnetar could act as a mechanism to achieve the exceptional luminosity of SLSN events (the internal engine model, e.g. \citealt{{Kasen2010}, {Dexter2013}}). These models have been proposed to explain the production of a reasonable subset of both the SLSN-II and SLSN-I events with moderately massive ($\sim$20-40 M$_{\odot}$) stars, and evidence suggests that engines are active in at least some SLSNe both  via detections of luminous X-rays \citep{Levan2013}, and detailed light curved modelling of SLSN events \citep{Inserra2013A,Nicholl2013,Nicholl2015A}. 

Clearly, the progenitors of SLSNe remain poorly understood, as the current lack of constraints upon the properties of the SNe explosions make it difficult to ascertain which progenitor models are correct. A powerful way of tackling this problem is to study the host galaxies of these extreme cosmic explosions, and infer progenitor properties from the environments in which they form. 
This method has been used effectively to constrain the properties of progenitors of other types of transient. For example, early differences between SNe Ia and SNe II could be inferred from the presence of the former in ancient elliptical galaxies, while the latter arise exclusively
in star forming hosts. More recently, increasingly sophisticated approaches have been made to study both the luminosities and morphologies of the host
galaxies of various transient types, along with their location within their hosts. Of particular relevance to SLSNe have been studies of the host
galaxies of long duration gamma-ray bursts (LGRBs). These events, the only stellar collapse events whose luminosities exceed those of SLSNe \citep{Bloom2009,Racusin2009}, have been shown to arise primarily from the brightest regions of low-mass mainly low metallicity hosts \citep[e.g.][]{{Fruchter2006},{Savaglio2009},{Svensson2010},{Perley2013}}. Such results imply that they arise from massive $>40$ M$_{\odot}$ low
metallicity stars \citep[e.g.][]{Fruchter2006,Larsson2007,Raskin2008,Graham2013}. Similar constraints have been derived for ``normal" SNe, suggestive
of an increasing mass spectrum from SN II $\rightarrow$ SN Ib $\rightarrow$ SN Ic \citep{James2006,Kelly2008}. 

There have been several studies of the host galaxies of SLSNe. \cite{Neill2011} found the host galaxies of SLSNe to be exceptionally faint and blue, compared to a sample of field galaxies, although this study was limited by the depth of the observations (GALEX and SDSS), with the majority of the more recently discovered, better characterised, but more distant SLSNe yielding only upper limits for their hosts in the UV and optical.  More recently \cite{Lunnan2014} carried out a survey of Type-I SLSN hosts, comparing the properties of their sample with those of the host galaxies of other core collapse events such as ``normal"  CCSNe and LGRBs. Their results implied that the hosts of SLSNe are less luminous and less metal 
rich than those of the general SNe population, but do exhibit comparable metallicities to the hosts of LGRBs, suggestive of similarities of progenitor 
between these two classes of event.  Alternatively, the study of Mg and Fe absorption lines in handful of SLSN hosts by \cite{Vreeswijk2014}, seems to suggest different progenitor paths for SLSN and LGRB events, due to the lower absorption strengths observed in SLSNe environments than in GRB hosts. 
A spectroscopic study of the hosts of SLSNe carried out by \cite{Leloudas2015} has shown the hosts of Type-I and Type-R events to possess extreme emission lines (Extreme Emission Line Galaxies), in contrast to SLSN-II hosts, which have comparatively softer radiation fields. The authors use this to support the notion of different progenitor systems for Type-I and Type-II events, advocating a massive, population III-like progenitor for H-poor SLSNe.  This is, however, contrary to the analysis of those who subscribe to a magnetar powered progenitor model \citep[e.g.][]{{Lunnan2014},{Inserra2013A}}, for which a slightly less massive progenitor \citep[$\textgreater$40M$_{\odot}$,][]{Davies2009} would suffice.   Although the samples presented by \cite{Lunnan2014} and \cite{Leloudas2015} have limited overlap, their distributions of metallicity are rather different, perhaps explaining the disparate conclusions. 

In some cases it is possible to directly study the immediate environments of the SNe, and determine the stellar populations at the explosion sites. Spectroscopic measurements of the local ($\sim$ kpc) environment of SLSN PTF12dam \citep{Thoene2014} have shown it to contain traces of recent star burst activity. Using the young stellar population and low metallicity of the region, the authors suggest a limit of $\textgreater$60M$_{\odot}$ upon the progenitor system, seemingly in agreement with low metallicity, massive population-III like progenitors inferred from global host properties.  

However, a study of the fractional host light contained within locations of Hydrogen-poor SLSNe within the ultra-violet carried out by \cite{Lunnan2015} (a method used to great effect with the host galaxies for other core collapse transients such as LGRBs and Type-Ic SNe to show a strong link between transient location and brightest star-forming regions within the host; \cite{{Fruchter2006},{Kelly2008},{Svensson2010},{Anderson2012}}) find that the locations of SLSNe-I are more concentrated on the light of their hosts than CCSNe, in which the probability of a CCSNe is roughly proportional to the surface brightness, but less concentrated than LGRBs. Given the strong link between stellar mass, stellar luminosity, and stellar lifetime this could naturally be explained by longer lived, possibly lower mass progenitors for SLSNe-I.

In this paper we present results from our survey of the hosts of SLSNe with the {\em Hubble Space Telescope} ({\em HST}) in the UV and nIR, complemented by a modest ground based programme of optical observations. These observations provide a view of the ongoing star formation via deep rest-frame UV observations, as well as a handle on any older populations within the hosts substantially expanding the wavelength baseline with respect to earlier surveys. In this paper we will focus on the broadband photometric properties of the host galaxies, demonstrating their origin in extremely small, low mass, and likely metal poor, systems. 

Throughout the paper we assume a standard $\Lambda$CDM cosmology with H$_{0}$=71 km s$^{-1}$ Mpc $^{-1}$ and $\Omega$$_{M}$=0.27 and $\Omega$$_{vac}$=0.73 \citep{Larson2011}. All reported magnitudes are given in the AB system and uncertainties are given at a 1$\sigma$ confidence level, unless otherwise stated.

\section{Sample and Observations}

Here we use nIR and rest-frame UV observations of a sample of 21 SLSN host galaxies, within a redshift range of $0.019 < z < 1.19$ (SN 2006gy $\rightarrow$ SCP 06F6). 

{\color{black} This HST sample (programme GO-13025; PI: Levan) comprised 21 targets, based on the sample of  \cite{Neill2011}, supplemented with luminous SNe from the literature (up to Jan 2012). This selection pre-dated more detailed sample work, such as that
by \cite{Gal-Yam2012} which introduced a cut at $M_V < -21$ for inclusion in an SLSNe sample. In particular, several of the original sample, while significantly more luminous
than typical SNe, were rather fainter than  $M_V < -21$, based on the reported magnitudes  and hence would be classed as luminous supernovae (LSNe) rather than
SLSNe. However, it should be noted that early examples such as SN~1995av, SN~1997cy and SN~2000ei have extremely limited
follow-up, and hence poorly know peak magnitudes, making their true nature uncertain. Conservatively we assign them as LSN in the absence of a detection of the SNe at a magnitude
of $M_V < -21$. Additionally, the nature of SN~1997cy remains debated, and it now seems likely that it is a Type Ia-SNe interacting with a hydrogen-rich shell of circumstellar material \citep[see][]{Hamuy2003}. Hence, we remove SN~1997cy from our sample of SLSNe for comparison with
other populations. Other SNe that do not make the peak-luminosity threshold for SLSNe are classified as ``LSN", while the unambiguous SLSNe sample is then used for our analysis and conclusions. This yields a sample of 17  SLSNe and 4 LSNe.  Unsurprisingly given the small contamination our conclusions are not significantly affected by the inclusion (or not) of LSNe). } 

\begin{figure}
	\centering
		\includegraphics[scale=0.45,angle=0]{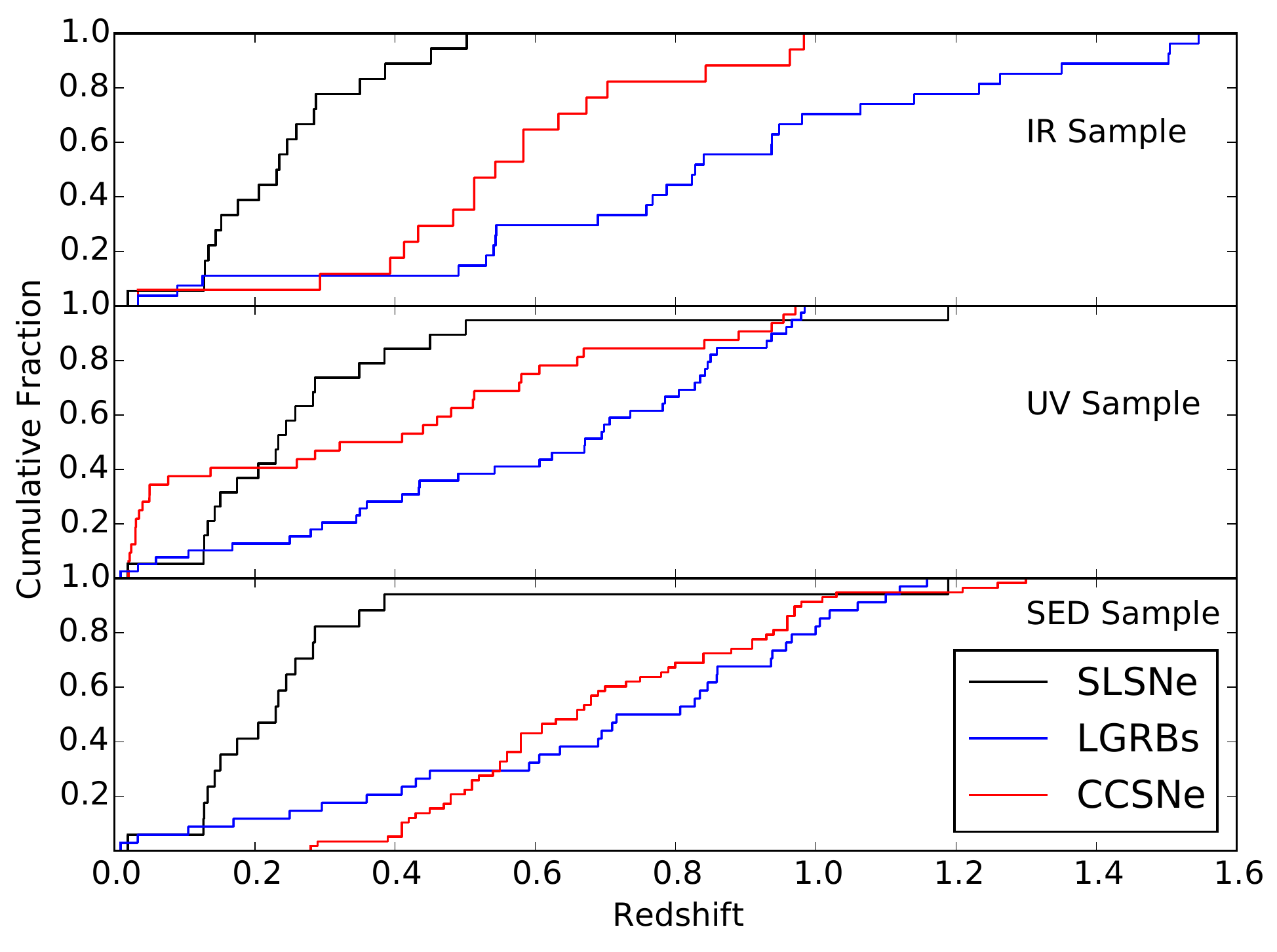}
	\caption{\color{black} {Redshift distribution of SLSN hosts used for comparisons in this work. We compare rest-frame IR (top) and UV (middle) 
	properties, as well as  masses and star formation rates
	derived from SED fitting (lower panel). Since these different diagnostics are available for only a fraction of each of our comparison samples the global redshift distribution is less appropriate. 
	Hence we show the redshift distribution for each sample separately. The SLSN host galaxies are typically at lower redshift $(z<0.5)$ than the GRBs, or than the GOODS CCSNe samples 
	to which we wish to compare. The possible impacts of this selection, and consideration of alternatives are presented in section 5.}}
	\label{fig:redshift}
\end{figure}

Table \ref{tab:targets} lists all the SNe targets used within this work and the distribution of redshifts for our sample is shown in Figure \ref{fig:redshift}.  Figure \ref{fig:redshift} also shows the redshift
distributions of host samples of core collapse SNe discovered in GOODS \citep[see also][]{Dahlen2003,Fruchter2006,Svensson2010} and of GRBs at {\color{black}$z\lesssim1.5$} \citep{Fruchter2006,Savaglio2009,Svensson2010} . We use these as a comparison sample of core collapse events that should represent both all core collapse systems creating a SNe (the GOODS CCSNe sample) and those occurring from only a restricted range of massive stars (probably those at low metallicity), represented by the GRBs.  We implement a redshift cut at {\color{black}$z\sim1.5$} on the GRB sample, in order to cover a comparable redshift range to our SLSNe, but not include the many high-z GRBs whose
host galaxies may differ because of the cosmological evolution of the galaxy population. 
It is now clear that low-z GRBs occur predominantly in smaller, lower-luminosity galaxies than the more distant bursts, probably due to their metallicity 
dependence, combined with the shifting mass-metallicity relation with redshift \citep{Perley2013,Perley2015b,Schulze2015}. Although this bias manifests itself predominantly below
$z \sim 1$ \citep{Perley2015b} it is possible that should SLSNe and LGRBs both exhibit metallicity bias, but at a different critical metallicity, then we could confuse 
evolution in galaxy properties with differing environmental constraints. Indeed, the survey of SLSN-I host galaxies reported by \cite{Lunnan2014} does
find some evidence for evolution, with lower-z SLSNe occurring in even smaller and lower luminosity galaxies. 
We note that {\color{black} within or pure SLSNe sample, 90\% of our SLSNe hosts lie at $z<0.4$}. Restricting our comparison samples
to these lower redshifts does not impact the nature of our conclusions, but given the much smaller sample sizes would impact the statistical significance.

\begin{table*}
\begin{minipage}{170mm}
\caption{SLSN host sample used within this paper, listing positions, redshifts and observations used within this study. Optical imaging with r' filter obtained with the William Herschel Telescope (WHT), B and R band imaging using the Keck I 10m Telescope and additional R band imaging obtained with Very Large Telescope (VLT) }
\label{tab:targets}
	\begin{tabular}{l l l l l l l l l l l l l l}
	 	\hline
		SLSN & Class 	& z 	& RA 	& Dec 	& UV 	& {\it{T}}$_{\text{exp}}$ & Optical  & {\it{T}}$_{\text{exp}}$ & nIR & {\it{T}}$_{\text{exp}}$ &   \color{black}{Ref.}\\
			 &		&	&(J2000)	&(J2000)	& Filter 	& UV (s) 	   	   & Band 	   &opt. (s) 	     	     &  Filter	     &nIR (s) 				&  \\  
 		\hline
		\hline
		SN1995av & \color{black}{LSN-IIn} & 0.300 & 02:01:41.34 & +03 39 38.9 & F390W & 1808 &  r$^\prime$ & 1500 & F160W &  206 & \color{black}{[1]} \\
		SN1997cy & \color{black}{Ia/IIn} & 0.063 & 04:32:54.86 & - 61:42:57.5 & F275W & 1832 & - & - & F160W & 206  &  \color{black}{[2]} \\
		SN1999as & \color{black}{SLSN-R} & 0.127 & 09:16:30.86 & +13:39:02.2  & F336W & 2032 &  - & - & F160W & 206  &  \color{black}{[3]} \\
		SN1999bd & \color{black}{SLSN-IIn} & 0.151 & 09:30:29.17 & +16:26:07.8  & F336W & 2036 & - & - & F160W & 206  &  \color{black}{[3]}\\
		SN2000ei & \color{black}{LSN-II} & 0.600 & 04:17:07.18 & +05:45:53.1  & F390W & 1808 & r$^\prime$ & 1500 & F160W & 206  &  \color{black}{[4]}\\
		SN2005ap &\color{black}{SLSN-I} & 0.283 & 13:01:14.83 &+27:43:31.4  & F390W & 1804 & R & 240 & F160W & 206 &  \color{black}{[5]} \\
		SN2006gy & \color{black}{SLSN-IIn} & 0.019 & 03:17:27.06 & +41:24:19.5 & F390W  & 932 & - & - & F160W & 206 &  \color{black}{[6,7]} \\
			     &    &          &		      &			    &	F275W & 846  &	&	& 	& 	&  & \\
		SN2006oz & \color{black}{SLSN-I} & 0.2860 & 22:08:53.56 & +00:53:50.4 & -&  - & r' & 300 & - & - &  \color{black}{[8]} \\
		SCP06F6 & \color{black}{SLSN-I} & 1.189 & 14:32:27.395 & +33:32:24.83 & F606W & 8054 & - & - & - & -&  \color{black}{[9]}\\
		SN2007bi & \color{black}{SLSN-R} & 0.128 & 13:19:20.19 & +08:55:44.3  & F336W & 1808 & - & - & F160W & 206  &  \color{black}{[10]} \\
		SN2008am & \color{black}{SLSN-IIn} & 0.234 & 12:28:36.30 & +15:34:50.0  & F336W & 1808 & - & - & F160W & 206 &  \color{black}{[11]} \\
		SN2008es & \color{black}{SLSN-II} &  0.202 & 11:56:49.13 & +54:27:25.7 & F336W & 1824 & r$^\prime$ & 1800 & F160W & 2812 &  \color{black}{[12, 13]} \\
					&	&		&			&		&	&		&F606W &	5630 &		&		&		   	 \\
					&	&		&			&		&	 &	 	& R		& 870	&		&		&	 \\
					&	&		&			&		&	 &	 	& B		& 900		&		&		&	 \\
		SN2008fz & \color{black}{SLSN-IIn} & 0.133 & 23:16:16.60 & +11:42:47.5 & F336W & 2032 & R & 1290 & F160W & 2612&  \color{black}{[14]}\\
			&	&		&			&		&	&	&	F606W &	5236 & 		&		&	&		 \\
			&	&		&			&		&	&	&	B & 1475	&	&	& & 	\\ 
		SN2009jh & \color{black}{SLSN-I} & 0.349 & 14:49:10.09 & +29:25:10.4 & F390W & 2044 & R & 240 & F160W &  2612   &  \color{black}{[15]}\\
			&	&		&			&		&	& 		&F606W &	5922 &			&			 \\
		PTF09atu & \color{black}{SLSN-I} & 0.501 & 16:30:24.55 & +23:38:25.0  & F390W & 2036 & r$^\prime$ & 1500 & F160W & 206  &  \color{black}{[15]}\\
		PTF09cnd & \color{black}{SLSN-I} &0.258 & 16:12:08.96 &+51:29:16.0-  & F390W & 2224 & r$^\prime$ & 1500 & F160W & 206 &  \color{black}{[15]}\\
		SN2010gx & \color{black}{SLSN-I} & 0.230 & 11:25:46.71 & - 08:49:41.4 & F390W & 1808 & - & - & F160W & 206 &  \color{black}{[8,15]} \\
		\color{black}{PTF10hgi} & \color{black}{LSN-I} & 0.10 & 16:37:47.00 & +06:12:32.3 &  -&  - & r$^\prime$ & 1500 & - & - &  \color{black}{[16]} \\
		PTF10vqv & \color{black}{SLSN-I} & 0.45 & 03:03:06.80 & -01:32:34.9 & -&  - & r$^\prime$ & 1500 & - & - &  \color{black}{[17]}\\
		SN2011kf & \color{black}{SLSN-I} & 0.245 & 14:36:57.53 & +16:30:56.7  & F336W & 2036 & - & - & F160W & 206  &  \color{black}{[16]}\\
		SN2011ke & \color{black}{SLSN-I} & 0.385 & 13:50:57.77 & +26:16:42.8 & F336W & 2044 & - & - & F160W & 206  &  \color{black}{[16]}\\
		PTF11dsf & \color{black}{SLSN-IIn} & 0.143 & 16:11:33.55 & +40:18:03.5 & F390W & 1832 & r$^\prime$ & 900 & F160W & 206 &  \color{black}{[18]} \\
		\color{black}{PTF11rks} & \color{black}{LSN-I} & 0.190 & 01:39:45.51 & +29:55:27.0  & F336W & 1804 & r$^\prime$ & 1800 & F160W & 206  &  \color{black}{[16]} \\
		SN2012il & \color{black}{SLSN-I} &  0.175 & 09:46:12.91 & +19:50:28.7 & F336W & 2036 & - & - &  F160W &  206  &  \color{black}{[16]}\\
				\hline
	\end{tabular}
	{{\color{black}References: [1] \cite{Richardson2002} (classification uncertain), [2] \cite{Hamuy2003}, [3] \cite{Gal-Yam2012}, [4] \cite{Schmidt2000}, [5] \cite{Quimby2007}, [6] \cite{Smith2007}, [7] \cite{Ofek2007}, [8] \cite{Leloudas2012}, [9] \cite{Barbary2009}, [10] \cite{Gal-Yam2009}, [11] \cite{Chatzopoulos2011}, [12] \cite{Miller2009}, [13] \cite{Gezari2009}, [14] \cite{Drake2010}, [15] \cite{Quimby2011a}, [16] \cite{Inserra2013A}, [17] \cite{Quimby2010b}, [18] \cite{Quimby2011b}. Note that PTF10hgi and PTF11rks are also sometimes referred to by their IAU designations of 
	SN 2010md and SN 2011kg respectively.}}
\end{minipage}
\end{table*}

\subsection{HST Data}\label{sect:hst-data}
We obtained \textit{HST} Wide Field Camera 3 (WFC3) images of the SLSN host galaxies for 21 hosts from of our sample. Observations were obtained in the rest-frame UV, probing the approximate rest-wavelength range of 2500-3500\AA~ and so we utilise F275W ($z<0.1$), F336W ($0.1 < z < 0.3$) and F390W ($0.3 < z < 0.6$) filters. In each orbit we also switched from the UV to nIR channel in WFC3 to enable us to obtain short nIR exposures ($\sim 200$s), which despite their duration are competitive with much longer ground based observations, reaching limits of H$_{AB}$ $\sim$ 25 (3$\sigma$). In addition, we obtained deep observations of the well studied
(and initially mysterious) SLSN, SCP 06F6 \citep{Barbary2009,Gaensicke2009} at $z=1.19$. For this event we obtained 3 orbits of exposure using ACS/WFC and F606W. A full log of observations is shown in Table~1. 

Some of the host galaxies were undetected in these exposures, suggesting extremely faint absolute magnitudes ($M_{nIR} >-15$). These were also targeted by a second programme (GO-13480; PI Levan), which obtained deeper optical observations using ACS in F606W and WFC3 again in the nIR. Again a full log is shown in Table~1, alongside additional optical observations of other SLSNe in our sample obtained from the ground with the William Herschel Telescope (WHT) and the Very Large Telescope (VLT). 

We stack and process our images within PyRAF using AstroDrizzle software \citep{Fruchter2002}. WFC3/UVIS and ACS images we drizzle
to a final pixel scale of 0.025\arcsec, while for the nIR images we retain the native 0.13\arcsec pixel scale due to the lack of dithering. Within the UV data set, images are subject to greater Charge Transfer Efficiency (CTE) losses, which arise due to inefficient transfer of charge between pixels during CCD readout, a consequence of cumulative radiation damage in a low Earth orbit environment \citep{Bourque2013}. To mitigate against this, all early images taken under programme GO-13025 utilised a pre-flash to fill charge traps, while in the latter observations we additionally moved the sources to the corners of the chip to minimise the number of transfers. The final individual images were then cleaned for CTE tails using the method of \cite{Anderson2010} prior to drizzling. 

The UV images were re-drizzled again to match the plate scale of the nIR imaging (0.13\arcsec pixel$^{-1}$). Though this lowers the resolution of the image, the technique allows for easier detection of low surface brightness features, and for a direct comparison between the nIR and UV imaging. 

Inclusive of our later, deeper imaging, we detected 18/21 of our HST imaged SLSNe in our rest frame UV imaging, and 19/20 in the nIR imaging. The hosts of some of the undetected SLSNe in our initial observations were recovered in the deeper exposures. Hence, we have host detections in at least a single band for $\sim$90\% of our \textit{HST} observed sample (all SLSNe excluding PTF09atu).

\subsection{Astrometry}
The majority of SLSNe from our sample possess discovery locations such that the SN position lies on, or close to, an underlying host detected within our HST imaging. Where possible, we perform initial astrometry measurements using discovery imaging where the SNe are as close as possible to maximum light. Imaging used in this procedure is described in table \ref{tab:discovery}. Astrometric measurements were carried out by aligning the discovery images by WCS for an initial approximation. Using routine IRAF tasks we determine the [x,y] centres of multiple matching sources in both discovery and HST fields, using point sources where available. Using IRAF tasks {\sc geomap} and {\sc geoxytran} we map and transform between coordinate systems for the two images, before transforming the [x,y] co-ordinates for the SN within the discovery image to the corresponding pixel within our HST imaging. This allows the SN position to be determined within the HST imaging. For four of our HST hosts for which discovery images were unavailable (namely SN~1997cy, SN~1999bd, SN~2000ei and SN~2011kf), we can only localise the SN position to the discovery RA and Dec, correcting for small offsets in HST's WCS solution by aligning it with 2MASS point sources.

We note that in the case of three SLSNe from our sample, initial astrometric measurements create some ambiguity in the identification of the real host. For SN~2000ei the presence of two galaxies within $\sim$ 1$\arcsec$ of the SN position precludes its unique identification. We test the chance probability of association (P$_{\text{chance}}$) that an unrelated galaxy of the same optical magnitude
or brighter would be found within the given offset from the apparent host for SN~2000ei from each of these nearby galaxies, using the method outlined within \cite{Bloom2002}. We adopt the host to the south west of the SN location, which has the lowest P$_{\text{chance}}$ value (=4.0x10$^{-3}$), as the true host to SN~2000ei.

Initial astrometric measurements for SN~2006gy suggest that the SN location is coincident with an unresolved ``knot'' of radiation approximately $\sim$ 1$\arcsec$ from the centre of NGC 1260, suggestive of perhaps a much smaller host satellite to the larger galaxy or that the SN continues to contribute strongly, even 8 years after the SN detection {\color{black} (see Figure \ref{fig:mosaic_06gy}). 
To test this we perform relative astrometry compared to an archival image of the SN, taken in November 2008 using the Wide Field and Planetary Camera 2 (WFPC2) in the F450W band (GO-10877, PI: Li). We find the SN position to be consistent with the centre of the source seen in our observations with a 0.08$\arcsec$ error circle. 

Subtraction of a point spread function reveals some possible features around the SN position, however, 
these could be faint features within the disc of NGC 1270, rather than extension of the source at the location of SN~2006gy. The source magnitude in our imaging of F390W(AB) = 22.6 $\pm$ 0.1 corresponds to an absolute magnitude of $\sim -11.7$, while if we assume that the source is unresolved (or at least the majority of the light arises from a very compact region) then the size is $<30$pc. This size is typical of a globular cluster, but the magnitude in blue light is too bright \citep[e.g.][]{Harris1996}. If it were a dwarf galaxy it would be relatively faint \citep[e.g.][]{Mcconnachie2012}, but unusually compact. In this case it may be an ultra-compact dwarf, a magnitude  fainter than the densest known example, M85-HCC1 \citep{Sandoval2015}, but comparable in size.
 Given the astrometric coincidence with the SN position it is then perhaps more likely the light continues to be dominated by SN emission (see also \cite{{Miller2010},{Fox2015}}), although in this case the minimal fading over the course of several thousand days is puzzling and also requires unusual explanations \citep{Fox2015}. Further observations will clearly be needed to distinguish between these possibilities. However, as the source is relatively faint, it does not significantly contribute to the photometric measurements of the host galaxy, and so does not impact our conclusions drawn for it.}

In the case of SN~2009jh, the SN apparently lies to the North-East of the host detected in deeper nIR imaging. We determine the P$_{\text{chance}}$ value of the apparent host of SN~2009jh, which we find to be 0.038 within an offset radius of 0.99$\arcsec$ (from the host half light radius and the SN's projected offset from the host centroid), indicating that for optical depth reached within our ACS imaging, the probability of the event being associated with another galaxy is low, but not especially so. Indeed, averaged over 20 hosts, we would expect a chance alignment with a sample of this brightness. We assign this nearby galaxy as the true host of SN~2009jh. 

We note that although the inclusion or exclusion of hosts SN~2006gy and SN~2009jh does not dramatically impact the results presented here, the host of SN~2006gy is the most luminous host in our sample by some margin, and so assigning it to a fainter satellite would result in some changes to the range of our distribution of SLSNe-II host luminosities.

\begin{figure*}
\begin{minipage}{180mm}
	\centering
		\includegraphics[scale=1.75,angle=0]{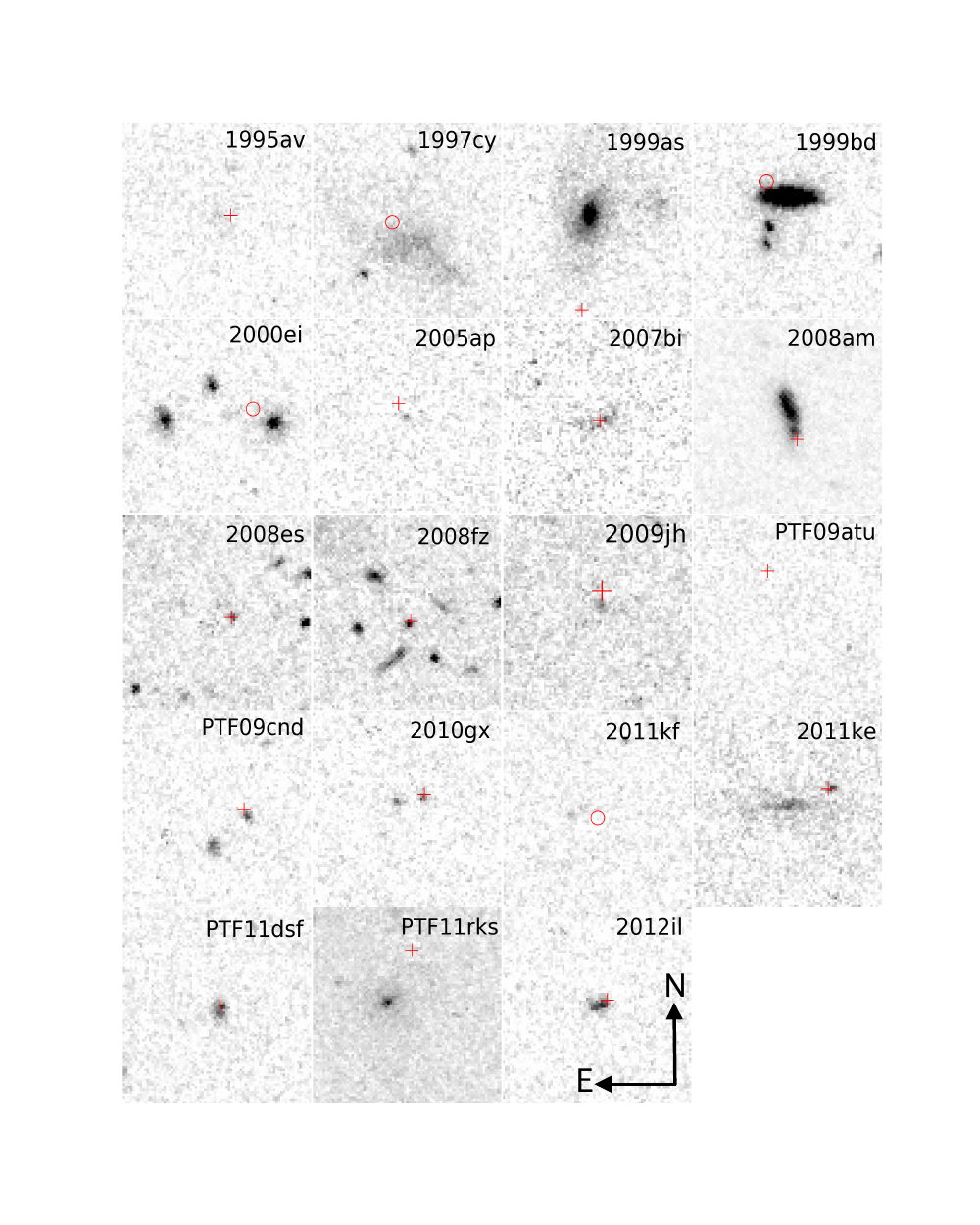}
	\caption{Host galaxies of SLSNe imaged in the nIR with \textit{HST} GO-13025 and GO-13480. Images  (bar SN 2006gy - see Figure \ref{fig:mosaic_06gy}) are scaled to 10$^{\prime\prime} \times $10$^{\prime\prime}$ and approximate SN positions are marked with red crosses where astrometry has been carried out, or circles located at the discovery coordinates of the SNe where discovery images were not available.}
	\label{fig:mosaic_ir}
\end{minipage}
\end{figure*}

\begin{figure*}
\begin{minipage}{180mm}
	\centering
		\includegraphics[scale=1.75,angle=0]{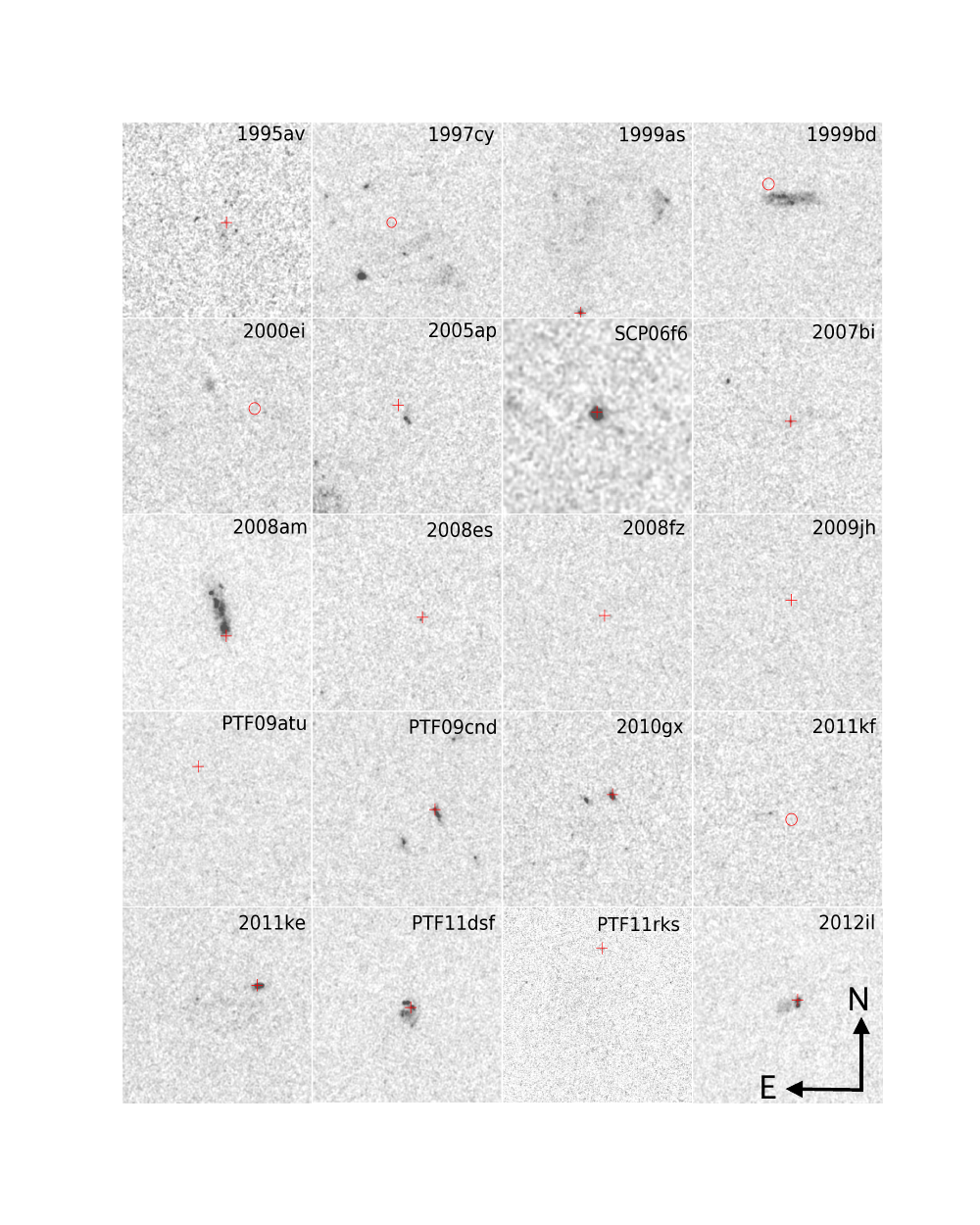}
	\caption{Host galaxies of SLSNe imaged in the UV with \textit{HST} GO-13025. Images (bar SN 2006gy - see Figure \ref{fig:mosaic_06gy}) are scaled to 10$^{\prime\prime} \times $10$^{\prime\prime}$ and drizzled to nIR pixel scale to highlight low surface brightness features. Approximate SN positions are marked with red crosses where astrometry has been carried out, circles located at the discovery coordinates of the SNe where discovery images were not available.}
	\label{fig:mosaic_uv}
\end{minipage}
\end{figure*}

A mosaic of our nIR and UV observations are shown in Figure~\ref{fig:mosaic_ir} and Figure~\ref{fig:mosaic_uv} respectively, (and our imaging of SN~2006gy in three bands shown within Figure \ref{fig:mosaic_06gy}) with the approximate location of the SN marked in each case (the detailed locations and statistics of the SLSNe positions within their hosts will be considered in forthcoming work, Angus et al. in prep). 

\begin{figure*}
\begin{minipage}{180mm}
	\centering
		\includegraphics[scale=0.85,angle=0]{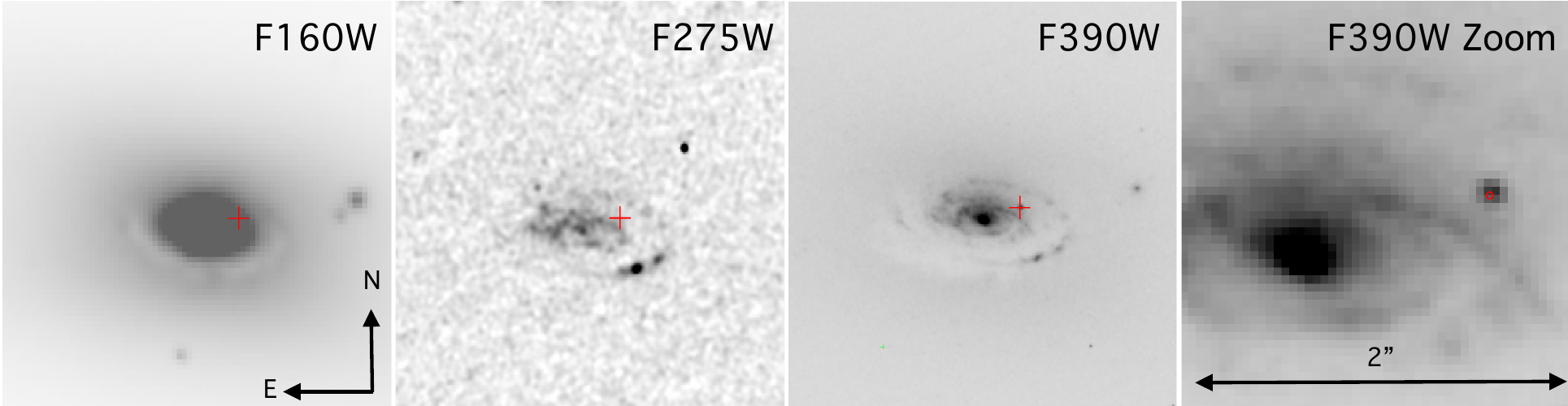}
	\caption{Host of SN~2006gy detected in F160W (first panel), F275W (second panel) and F390W (third panel). Images are scaled to 10$^{\prime\prime} \times $10$^{\prime\prime}$. The SNe location, {\color{black} determined from astrometric measurements from late time imaging with WFPC2, is marked in red}. We draw attention to a possible satellite to the larger host, coincident with the SNe location, revealed within our F390W imaging. To highlight this we provide a 2 '' x 2 '' image zooming in on this region and the central bulge in the fourth panel, where the SNe position is marked by a {\color{black}0.08"} error circle }
	\label{fig:mosaic_06gy}
\end{minipage}
\end{figure*}

\subsection{Ground-based Observations}
We supplement our HST observations with our own various ground based programmes. We carried out a service mode programme using ACAM on the William Herschel Telescope to obtain relatively shallow optical imaging ($\sim$1500 - 1800s of sdss r$^\prime$ band) for a small sample of SLSN hosts (namely SN~1995av, SN~2000ei, SN~2006oz, SN~2008es, PTF09atu, PTF09cnd, { \color{black}PTF10hgi}, PTF10vqv, PTF11dsf and {\color{black}PTF11rks}). 

We acquired R-band imaging of three galaxies from our sample (namely the hosts of SN~2005ap, SN~2008fz and SN~2009jh) using the FOcal Reducer and low dispersion Spectrograph \citep[FORS2;][]{FORS} on the VLT during the nights of 2013-08-31, 2014-01-24 and 2014-02-01. We obtained 239 seconds of $R$-band imaging for each host, reducing the images using standard procedures within IRAF. We recover faint unresolved detections of each host galaxy within our imaging. The calculated aperture magnitudes obtained for our SED fitting are provided within Table \ref{table:slsn_magnitudes}. 

We acquired deep optical imaging of the two faintest targets in the low-$z$ sample (SN 2008es and SN 2008fz) using the Low Resolution Imaging Spectrometer \citep{LRIS} on the the Keck I 10m telescope, during the night of 2013-12-04.  We obtained 900 seconds of $B$-band imaging and 870 seconds of $R$-band imaging of the host of SN 2008es, and 1475 seconds of $B$-band imaging and 1290 seconds of $R$-band imaging of the host of SN 2008fz.  Images were reduced using standard procedures via an automated pipeline (LPipe).  We recover faint, unresolved detections of the host galaxy of both targets in both filters, and calculate aperture magnitudes for our SED analysis.

\section{Photometry} \label{sect:phot}
Photometry was performed using one of two different methods, dependant upon the appearance of the host galaxy within the HST imaging. Where possible we used the automatic detection and extraction package, Source Extractor \citep[here after SExtractor;][]{Bertin1996}, for which the program parameters were adjusted accordingly for each host to optimise its detection and extraction. Here we applied a surface brightness signal to noise cut of cut of two per pixel for nIR images and one for UV images, in order to include faint surface brightness features.  We report measured host magnitudes as {\tt MAG\_AUTO}
values which attempt to account for additional light outside of the notional aperture. Several galaxies show a light distribution dominated by individual bright knots in ultraviolet imaging, and the deblending parameters were adjusted for each host to ensure it was not broken into multiple components. For this work the nIR images were used to determine which UV components should be included in the analysis, as these bands are dominated by a smoother light profile arising from older stars. Zeropoints for each filter were taken from the STScI WFC3 handbook \citep{WFC3}.

We also utilise straightforward aperture photometry, setting large apertures to encompass the majority of the light of the galaxy, and determining the background via the use of a large number ($>20$) of sky apertures. This technique gave results consistent with those determined via SExtractor, and was used to obtain $3 \sigma$ limits where necessary. In these cases we apply an aperture correction determined by the estimated encircled energy curves of WFC3 detectors \citep{WFC3}. In the case of a detection in one band but no detection in another, the size of the aperture used to determine the upper limit was set equal to that used to measure the magnitude in the band where the source was detected. 

Additional photometry of hosts imaged using the WHT and VLT in r' and Johnson-Morgan R bands respectively was carried out in a like manner to the HST images, applying a  surface signal to noise cut off of one per pixel before extraction. Photometry of galaxies on ground based images was carried out relative to SDSS observations of the same field, and is given in the r' band. We correct all our photometry for Galactic extinction using the Milky Way dust maps of \cite{Schlafly2011} (via the NASA/IPAC Infrared Science Archive\footnote{http://irsa.ipac.caltech.edu/applications/DUST/}) for the appropriate image filter.

\subsection{Comparison Samples}\label{sect:compare}
While the properties of the SLSN hosts themselves are of interest, they are most diagnostic when compared to other classes of extragalactic transient whose progenitors
are better understood. To this end, we employ a comparison sample of LGRB and CCSN\footnote{Here, we use CCSNe to define all core collapse events, including SN Ib, Ic, II and their various sub-types. Where appropriate and possible, we specify the SN type} host galaxies. In principle, CCSNe  should trace all core collapse events, although the mass function means they will be dominated by stars at the lower mass end ($\sim 8$ M$_{\odot}$ to $\sim25$ M$_{\odot}$). There also remains a possibility that some very massive stars can undergo core collapse without yielding a luminous supernovae \citep[e.g.][]{Smartt2009,Ugliano2012,Kochanek2014} such that CCSNe samples might only provide a census of lower mass core collapsing stars (e.g. 8$\textless$ M$_{*}\textless$ 20 M$_{\odot}$). Indeed, constraints from explosion parameters have shown the majority of CCSNe to be consistent with lower mass progenitors, as opposed to more massive Wolf Rayet stars \citep{Cano2013,Lyman2014} GRBs likely represent a population with rather larger initial masses \citep{Larsson2007,Raskin2008}.  LGRBs are now known to be associated with the core collapse of massive stars, and broad line SN Ic are near ubiquitously associated with low-z events \citep[where such signatures can be seen,][]{Hjorth2012}. When compared to the hosts of CCSNe they are generally smaller and of lower luminosity, consistent with an origin in galaxies of lower metallicity \citep{Fruchter2006,Svensson2010}. In relatively local examples, where spatially resolved gas phase metallicities can be obtained, these indeed appear to be lower for GRBs than for CCSNe, even in cases where the luminosity of the galaxy is relatively high (i.e. the GRB host galaxies lie off the mass-metallicity relation, \citealt{Modjaz2008,Graham2013}). Hence, comparing the hosts of SLSNe to these events allows us to test the large scale environments of SLSNe against those of the bulk core collapse population and a subset which appears to derive largely from massive stars at lower metallicity, although we note that agreement on this matter is not complete \citep[e.g. ][]{Podsiadlowski1992,Eldridge2008,Smartt2009,Drout2011}.  By exploiting both LGRB and CCSN host samples we can ascertain if there is a strong metallicity dependence in SLSN production, and if this is more or less extreme than that observed in GRB hosts. 

The observed samples are undoubtedly biased against highly dusty lines of sight such that the most dusty examples are missed. This effect has been well studied in GRBs \citep[e.g.][]{Jakobsson2006,Fynbo2009}, and the inclusion of dusty sight lines does apparently extend the GRB host mass function to higher masses than if they are excluded \citep[e.g.][]{Perley2013}. However, the effect below $z \sim 1.2$, where our comparisons to SLSNe are 
conducted, is small, with very few dusty massive systems \citep[\citealt{Perley2015a,Perley2015b}, although see][]{Stanway2015}. The impact on 
SNe detection may be even larger given their fainter peak magnitudes and uniquely optical selection.  

Tables \ref{table:IR_CCSNe_Comp}, \ref{table:IR_LGRB_Comp} and \ref{table:UV_Comp} list the names, locations and redshifts of the host galaxies for direct photometric comparison. We make our own photometric measurements for hosts with available HST rest frame UV or nIR imaging, and draw from literature elsewhere.

Our LGRB host sample contains events at $z\lesssim1.2$ (for broad matching of the SLSNe redshift distribution, and comparable sample size). Rest frame UV observations
are obtained from the literature \citep[in particular utilising the GHostS project][for other references see Table \ref{table:UV_Comp}]{Savaglio2009}. nIR observations are obtained from GHosts, and also from our {\em HST} snapshot programme GO-12307; PI Levan, Lyman et al. in prep).

Our CCSNe host sample is based on that detected in the rolling SNe searches of the GOODS field \citep{Dahlen2003,Fruchter2006,Riess2007,Svensson2010,Strolger2010}. These tiled the GOODS field repeatedly in the F850LP filter, with a cadence of $\sim 45$ days, primarily chosen to locate SNe Ia at $z>1$. However, this search also provides an untargeted and highly sensitive moderate redshift ($0.1 < z \lesssim1$) survey for core-collapse events. Subsequently, the GOODS field has been observed in the nIR with both NICMOS and WFC3, and more recently in the blue using ACS and WFC. We use these images to obtain nIR magnitudes for the CCSN hosts, and for rest-frame UV magnitudes where field coverage and redshifts allow, performing photometry as described above for the SLSNe population.

Due to restrictions in field coverage and probed rest-frame wavelength from the GOODS UV field imaging, we supplement our CCSNe host comparison sample with that of \cite{Sanders2012}, which provides an untargeted, albeit typically low redshift, sample of stripped envelope SNe hosts. For these hosts we draw upon literature values to determine their rest-frame UV brightness.  

\section{Determining the Physical Parameters}\label{sect:properties}

The redshifts of all of our sample of both SLSNe and comparison objects are known, and hence we can compare the physical properties of the galaxies. Of particular use can be a simple comparison of observed properties to physical properties over a similar redshift range, especially in cases for which Spectral Energy Distribution coverage is poor. In particular, we can compare the absolute magnitudes at UV and nIR wavelengths, using these as proxies for star formation rate and stellar mass respectively. 

We apply Spectral Energy Distribution fitting to {\color{black} all of} our hosts, to constrain masses, ages and star formation rates. We also measure the sizes of the host galaxies, specifically the radius within which 80\% of their light is contained (following \citealt{Fruchter2006,Svensson2010}). 
 
\subsection{Spectral Energy Distribution Fitting}

We have photometry from rest-frame nIR to near UV in all cases, with an extension to the mid-IR for brighter hosts, which allows us to fit template spectral energy distributions. To do this we supplement our own photometric measurements with those from other public data and literature. We use Sloan Digital Sky Survey (SDSS) images \citep{SDSS2012} to extract optical photometry for our hosts using the same techniques applied to our own ground based imaging, in some cases this is supplemented with observations with Catalina Real-time Transient Survey \citep[CRTS;][]{CRTS2009}, and additionally use mid-IR observations from WISE \citep{WISE2013}. Finally, we also utilise published photometry of individual SLSN host galaxies from \cite{Germany2000,Quimby2007,Barbary2009, Neill2011, Hudelot2012, Leloudas2012} and \cite{Lunnan2014} to complete our SEDs. {\color{black}For all photometry, we utilise the MAG\_AUTO function within SExtractor, which models and accounts through fitting Kron-like elliptical apertures to the source, in order to minimise any differences in the fraction of host light across different bands.}

The acquisition of both nIR and UV data points allows us to simultaneously fit both masses and star formation rates, which when combined with the depth of the imaging provides better constraints upon the blue and red ends of spectra when fitting, achieving more realistic estimates of host properties than previous SED fitting attempts.
The broad-band observations are fitted against the template model chosen to derive masses, ages and star formation rates for these hosts, a more detailed outline of which can be found within \cite{Perley2013}. Here we assume a mass-dependent metallicity and a host ionization parameter {\color{black}of 4$\times$10$^{7}$, except in the case of SN~2011kf and SN~2011ke, as in both of these  SED's there was an observable excess of flux within the filters corresponding to rest-frame O[III] or H$\alpha$ lines when compared to a fit with no nebular emission. In these cases we draw this parameter from the previous spectroscopic studies of \citealt[][]{Lunnan2014} and \cite{Leloudas2015}.}  

\subsection{Luminosity Diagnostics}

Whilst SED fits allow us to determine the properties of host galaxies to a relatively high degree of precision, the constraints of an SED fit are strongly dependent upon the number and wavelength range of bands used to fit the template spectra. The properties derived are also highly sensitive to star formation history adopted during the fitting procedure. For simplicity, and for direct comparison with previous work we therefore also consider nIR and UV rest frame luminosities as direct proxies for the stellar mass and star formation rate\footnote{In figures in which the main x-axis shows an observed absolute magnitude, the upper axis therefore shows the mass/star formation rate inferred from these proxies, while figures showing physical parameters are those derived from SED fits}. To do this we utilize the relations used in \cite{Savaglio2009} for stellar
mass, namely that 

\begin{equation}
	\log{M_{*}} = -0.467 M_{nIR} - 0.179
\end{equation}

We can also directly convert our rest frame UV luminosities into star formation rates as per \cite{Kennicutt1998};
\begin{equation}
	SFR(\mathrm{M}_{\odot}~\mathrm{yr}^{-1}) = 1.4 \times 10^{-28} L_{\nu}
\end{equation}
where L$_{\nu}$ is in cgs units of erg s$^{-1}$ Hz$^{-1}$ in the rest-frame wavelength range from 2500-3500\AA, a region in which all our
UV observations lie. This relation assumes a constant star formation over a 100 Myr period with a specified initial mass function. Utilizing both mass and SFR we can also calculate a specific star formation rate, $ \Phi = \frac{SFR}{M_{*}}$. These values
generally give results comparable to those from our direct SED fitting. 

Finally, in addition to straight forward photometry, SExtractor also can be used to ascertain the fractional light radii of host galaxies using the {\tt FRAC$\_$LIGHT} parameter \citep{Bertin1996}, which fits an isophotal profile to a source then measures the relative size of the source in pixels, later converted into kiloparsecs using the plate scale. We use the common LGRB host diagnostic of radii containing 80\% of the total flux from the host (r$_{80}$) within the nIR F160W images.

Errors for the SExtractor measurement of r$_{80}$ in pixels for the hosts were estimated by modelling the capability of SExtractor to detect the full radial profile of a source at given magnitude and redshift as a function the image noise. An artificial field of objects were generated using IRAF routines, with artificial galaxies specified to span a similar apparent magnitude and surface brightness range to our host galaxies in the F160W band. The discrepancy between specified object size and that measured by SExtractor was measured, with different levels of simulated noise, suggesting that significant errors can arise for sources close to the noise limit. These errors are provided in Table~{\ref{table:slsn_magnitudes}}, alongside our r$_{80}$ estimates.

\begin{landscape}
\begingroup
\begin{table}
\begin{minipage}{250mm}
\centering
\caption{Apparent and absolute magnitudes of SLSN hosts observed with HST in nIR (F160W), rest-frame UV (F275W,F336W or F390W) and optical bands. Optical photometric properties of a subset of this sample as observed with $^a$ WHT (r'), $^b$ VLT (R band) and $^c$ HST ACS (F606W). We also present apparent r$_{80}$ sizes of our HST SLSN hosts as detected within WFC3 F160W imaging. Optical imaging key: $^{a}$ r$^{\prime}$ band, $^{b}$ R band, $^{c}$ B band.}
\label{table:slsn_magnitudes}
	\begin{tabular}{l l l l l l l l l l l l l}
		\hline
		SLSN & m$_{UV}$ & M$_{UV}$  & m$_{optical}$& M$_{optical}$ &  m$_{nIR} $ & M$_{nIR}$  & r$_{80}$  &r$_{80}$  \\
			& AB mag & AB mag	&AB mag		&	AB mag	&	AB mag	& AB mag	&	(kpc) 	& (arcsec) \\	
		\hline
		\hline
		SN1995av		& 24.97 	$\pm$ 0.32 		& -15.82 $\pm$ 0.32 		& 23.77 $\pm$ 0.19 $^{a}$ 	& -16.96  $\pm$ 0.19 $^{a}$	& 23.17 $\pm$ 0.27 		& -17.51 $\pm$ 0.27  	& 10.66 $\pm$ 3.07 & 2.41 $\pm$ 0.69 \\
		SN1997cy		& 21.14 	$\pm$ 0.21 		& -16.17 $\pm$ 0.21 		& - 						& - 	 		 			& 20.19  $\pm$ 0.03 		& -16.98 $\pm$ 0.03  	& 3.29  $\pm$ 1.10 & 2.75 $\pm$ 0.91  \\
		SN1999as		& 21.15 	$\pm$ 0.10 		& -17.70 $\pm$ 0.10 		& - 						& - 	 		 			& 19.19 $\pm$ 0.03 		& -19.55 $\pm$ 0.03  	& 6.81  $\pm$ 2.39 & 3.03 $\pm$ 1.07  \\
		SN1999bd		& 21.85 	$\pm$ 0.06 		& -17.40 $\pm$ 0.06 		& - 						& - 	 		 			& 18.779 $\pm$ 0.003 	& -20.349 $\pm$ 0.003 	& 2.95  $\pm$ 1.02 & 1.13 $\pm$ 0.39  \\
		SN2000ei 		& 23.81 	$\pm$ 0.21 		& -19.29 $\pm$ 0.21 		& 22.67 $\pm$ 0.14 $^{a}$ 	& -20.07  $\pm$ 0.14 $^{a}$	& 20.90 $\pm$ 0.03 		& -21.44 $\pm$ 0.03  	& 6.55  $\pm$ 2.00 & 0.98 $\pm$ 0.30  \\
		SN2005ap 		& 24.32 	$\pm$ 0.09 		& -16.24 $\pm$ 0.09 		& 23.64 $\pm$ 0.27 $^{a}$ 	& -16.90  $\pm$ 0.27 $^{a}$	& 23.48  $\pm$ 0.36 		& -17.05 $\pm$ 0.36  	& 3.23  $\pm$ 0.85 & 0.76  $\pm$ 0.20  \\
		SN2006gy 		& 19.86 	$\pm$ 0.01 		& -15.55 $\pm$ 0.01 		& - 			 			& - 	 					& 11.951 $\pm$ 0.001 	& -22.661 $\pm$ 0.001 	& 3.69  $\pm$ 2.11 & 9.70 $\pm$ 5.56  \\
		SN2006oz 		& - 		 				& - 					& 24.09 $\pm$ 0.26 $^{a}$  	& -16.56  $\pm$ 0.26 $^{a}$	& - 					& - 	  			  	& -	 			   & -  			 \\
		SCP06F6 	& 27.88 	$\pm$ 0.20 			& -15.88 $\pm$ 0.20 		& - 						& - 	 					& -						& -   	   			  	& -      			   & -   			  \\
		SN2007bi 		& 23.83 	$\pm$ 0.28 		& -15.03 $\pm$ 0.28 		& - 						& - 	 					& 22.07 $\pm$ 0.18 		& -16.68 $\pm$ 0.18  	& 2.62  $\pm$ 0.77 & 1.16  $\pm$ 0.34  \\
		SN2008am 		& 21.20 	$\pm$ 0.026 		& -19.00 $\pm$ 0.026	& - 						& - 	 					& 19.48 $\pm$ 0.006 	& -20.63 $\pm$ 0.006 	& 4.31  $\pm$ 1.44 & 1.17  $\pm$ 0.39  \\
		SN2008es 		& $\textgreater$25.32 		& $\textgreater$-14.526 	& 25.96 $\pm$ 0.20 $^{b}$ 	& -13.86  $\pm$ 0.20 $^{b}$	& 26.85 $\pm$ 0.40 		& -12.95 $\pm$ 0.40  	& 1.19  $\pm$ 0.23 & 0.36 $\pm$ 0.07  \\
					&  						& 					& 26.96 $\pm$ 0.25 $^{c}$ 	&  -12.85 $\pm$ 0.25 $^{c}$ 			& 			 		& 				  	&							  \\
		
		SN2008fz 		& 26.73 	$\pm$ 0.55		& -12.28 $\pm$ 0.55 		& 25.58 $\pm$ 0.19 $^{b}$	& -13.33 $\pm$ 0.19 $^{b}$	& 25.18 $\pm$ 0.06 		& -13.66 $\pm$ 0.06  	& 0.87  $\pm$ 0.18 & 0.37 $\pm$ 0.08  \\
			 		&  						& 					& 26.17 $\pm$ 0.22  $^{c}$ 	&  -12.81 $\pm$  0.22$^{c}$ 	& 			 		& 				  	&							  \\
		SN2009jh 		& $\textgreater$25.92		& $\textgreater$-15.139 	& 25.46 $\pm$ 0.07 $^{b}$	& -15.59  $\pm$ 0.07 $^{b}$	& 25.30 $\pm$ 0.15 		& -15.71 $\pm$ 0.15  	& 2.71 $\pm$ 0.63 & 0.55 $\pm$ 0.13     				 \\
		2010gx 		& 23.96 	$\pm$ 0.04 		& -16.24 $\pm$ 0.04 		& - 						& - 	 		 			& 23.17 $\pm$ 0.15 		& -16.90 $\pm$ 0.15  	& 1.84  $\pm$ 0.46 & 0.51 $\pm$ 0.13  \\
		PTF09atu 		& $\textgreater$25.47 		& $\textgreater$-16.533 	& $\textgreater$23.14 $^{a}$ 		& $\textgreater$-18.79  	$^{a}$	& $\textgreater$23.39 	&$\textgreater$-18.452	& -  			   & -   			  \\
		PTF09cnd 		& 24.01 	$\pm$ 0.05 		& -16.40 $\pm$ 0.05 		& 23.60 $\pm$ 0.04 $^{a}$	& -16.768  $\pm$ 0.04 $^{a}$	& 22.56 $\pm$ 0.12 		& -17.76 $\pm$ 0.12  	& 3.11  $\pm$ 0.89 & 0.78 $\pm$ 0.22  \\
		 \color{black}{PTF10hgi} 		& - 						& - 	 				& 22.05 $\pm$ 0.06 $^{a}$	& -16.329  $\pm$ 0.06 $^{a}$	& - 					& - 				  	& - 	  		   & - 				 \\
		PTF10vqv 		& - 						& - 					& 23.33 $\pm$ 0.12 $^{a}$	& -18.392  $\pm$ 0.12 $^{a}$	& - 	 				& - 	  			  	& - 	  		   & - 	   			 \\
		2011kf 		& 24.51 	$\pm$ 0.38 		& -15.78 $\pm$ 0.38 		& - 						& - 	 					& 24.06 $\pm$ 0.40 		& -16.14 $\pm$ 0.40  	& 1.24  $\pm$ 0.27 & 0.33 $\pm$ 0.07  \\
		2011ke 		& 23.12 	$\pm$ 0.03 		& -15.92 $\pm$ 0.03 		& - 						& - 	 					& 23.21 $\pm$ 0.14 		& -15.78 $\pm$ 0.14  	& 5.17  $\pm$ 1.48 & 2.08 $\pm$ 0.60  \\
		PTF11dsf 		& 22.88 	$\pm$ 0.04 		& -18.38  $\pm$ 0.04 	& 22.04 $\pm$ 0.11 $^{a}$	& -19.204  $\pm$ 0.11 $^{a}$	& 21.81 $\pm$ 0.07 		& -19.41 $\pm$ 0.07  	& 3.48  $\pm$ 0.98 & 0.67 $\pm$ 0.19  \\
		 \color{black}{PTF11rks} 		& 22.43 	$\pm$ 0.16 		& -17.38 $\pm$ 0.16 		& 20.95 $\pm$ 0.25 $^{a}$ 	& -18.77  $\pm$ 0.25 $^{a}$	& 20.69 $\pm$ 0.06		& -18.96 $\pm$ 0.06  	& 5.55  $\pm$ 1.77 & 1.77 $\pm$ 0.57  \\
		SN2012il 		& 22.78 	$\pm$ 0.06 		& -16.75 $\pm$ 0.06 		& - 						& - 	 					& 21.82 $\pm$ 0.06 		& -17.63 $\pm$ 0.06  	& 1.89  $\pm$ 0.53 & 0.64 $\pm$ 0.18  \\
		\hline 
 	\end{tabular}
\end{minipage}
\end{table}
\endgroup
\end{landscape}

\section{Results}
Below we first present the measured nIR and UV luminosities of the host galaxies, and consider the implications these results have when treated as proxies for stellar mass and SFR, respectively, before evaluating the derived SED properties of our hosts, and their physical sizes. {\color{black} We then compare these to our comparison samples. In the majority of cases
it is apparent that the SLSNe hosts bear little similarity with any other core collapse host population, being both fainter and smaller, we consider the implications of this in section 6.} 

The photometric UV, optical and nIR magnitudes of the {\color{black}LSN and} SLSN host sample considered within this work are presented in Table \ref{table:slsn_magnitudes}, and the derived UV and nIR host properties from SED fitting of these hosts are presented in Table \ref{table:slsn_properties}. The direct photometric measurements and derived properties of our chosen LGRB and CCSN comparison samples are presented for nIR and UV observations within Tables \ref{table:IR_CCSNe_Comp}, \ref{table:IR_LGRB_Comp} and \ref{table:UV_Comp} respectively.

\begin{figure}
	\centering
		\includegraphics[scale=0.45,angle=0]{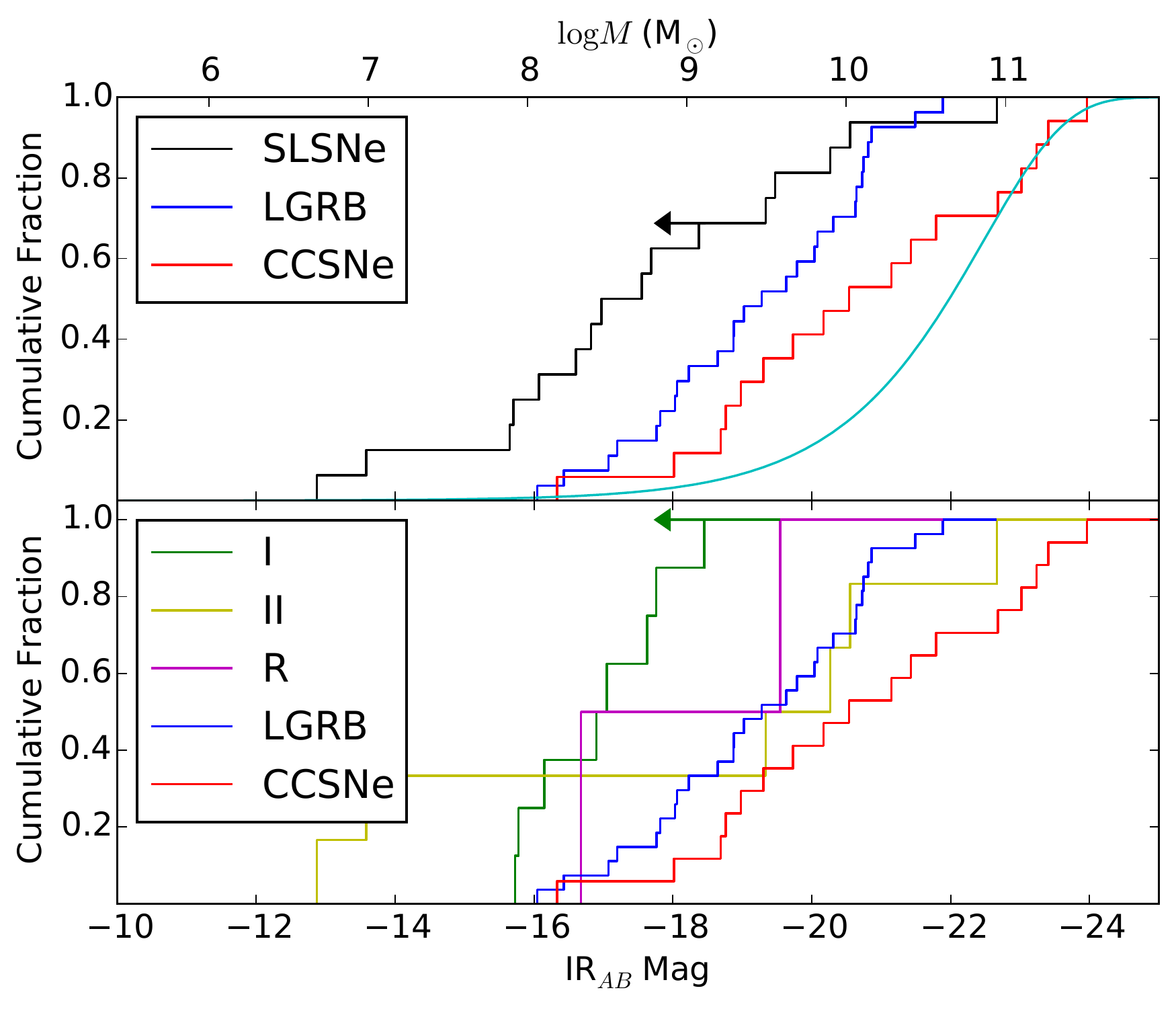}
	\caption{Upper panel: Cumulative frequency distribution of the absolute nIR magnitudes of core collapse event host galaxies. Arrows represent cases in which no host was detected and the 3$\sigma$ limiting magnitude is used to place an upper limit upon the brightness of these hosts. The difference between the distributions of the SLSN and other core collapse hosts is statistically significant, with probabilities of {\color{black}0.008} and {\color{black}0.0017} of the SLSN hosts being drawn from the same population as LGRB and CCSN hosts respectively. We also display the NIR galaxy luminosity function for galaxies within our brightness range (cyan line) \citep{Cirasuolo2007}. Using nIR brightness as a proxy for mass (top x-axis), we can expect our hosts to be significantly less massive too.
	Lower panel: we present the same distributions with the hosts of SLSNe broken down by classification.  Here the SLSN classes appear indistinguishable from one another in brightness, but this is likely due to small number statistics. We perform AD testing between the different subclasses and both core collapse comparison groups, and find SLSN-I hosts to be inconsistent with our sample of core collapse transients, although we find a stronger association for SLSN-II, due to the much broader distribution in brightness it exhibits.}
	\label{fig:cdf_ir}
\end{figure}

In Figure \ref{fig:cdf_ir}, we present the cumulative distribution of the absolute nIR magnitudes of the SLSNe hosts against those of the LGRBs and a subsample of GOODS CCSN hosts for which parallel photometric measurements were carried out. It can be seen here that SLSN hosts are in most cases much fainter than either LGRB hosts or CCSN hosts over the redshift range considered here. Breaking down by SLSN sub-type, the most extreme examples (ignoring the small sample size of SLSNe-R) are the SLSN-I hosts, which are inconsistent with any other population of transient hosts. In contrast the SLSN-II hosts extend to magnitudes much fainter than CCSN host galaxies
but at the brighter end of their distribution are comparable to the luminosities of LGRB hosts. In addition to the observed populations we also show
as a solid cyan line the expected distribution of host magnitudes should they be drawn from the field population in proportion to the total nIR luminosity density (i.e. uniformly from the luminosity weighted luminosity function, \citealt[][]{nIR_Lum}), demonstrating that all transient types arise from fainter galaxies than expected in this scenario. This is not surprising since weighting the luminosity function by the nIR is approximately equivalent to weighting by galaxy mass, and as such we see a significant contribution from massive, but largely quiescent galaxies which will not host core collapse events.

\begin{figure}
\centering
\includegraphics[scale=0.45,angle=0]{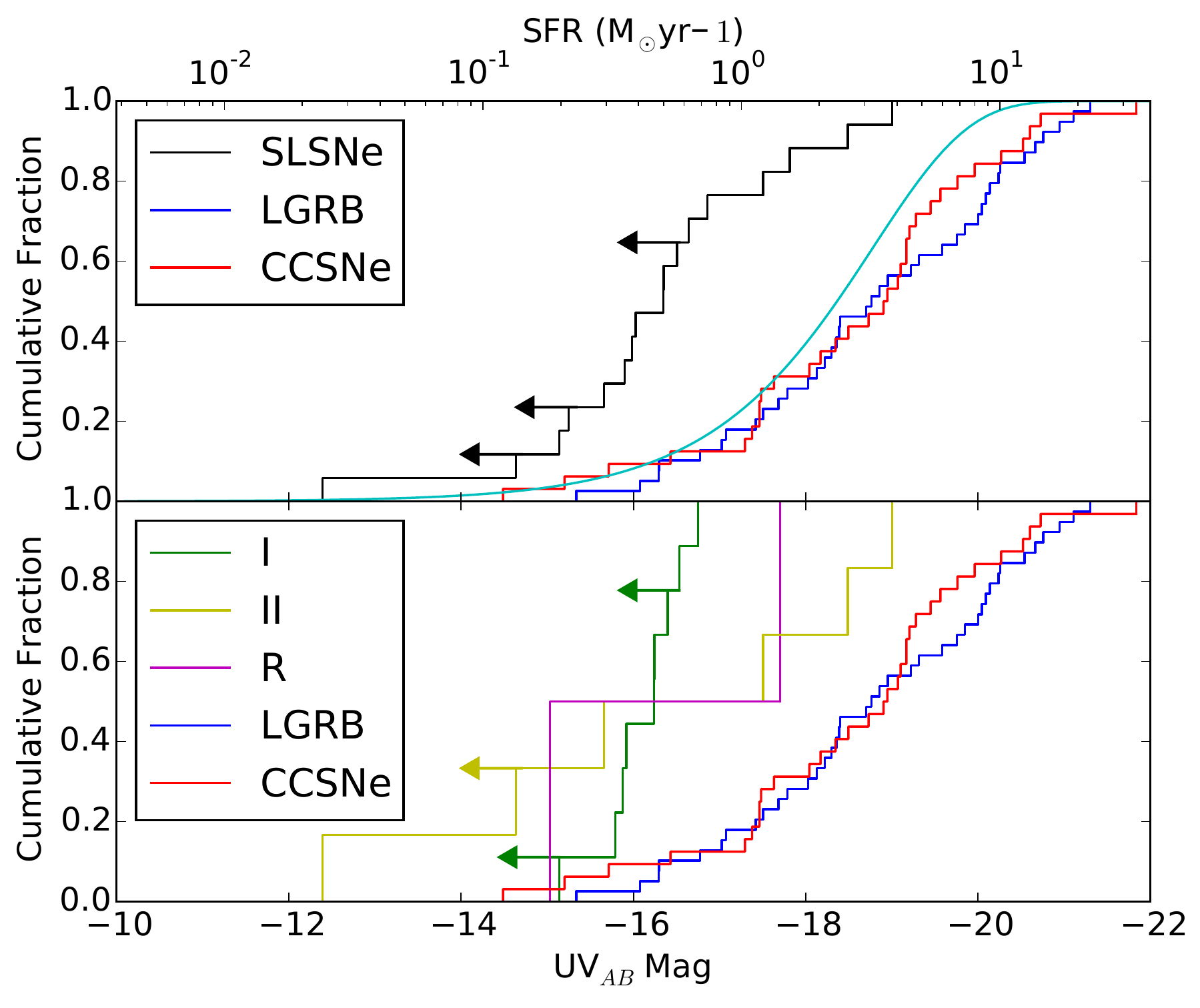}
\caption{Cumulative distribution of the UV luminosities of SLSN, LGRB and CCSN host galaxies (upper panel). Anderson-Darling results show that SLSN hosts are not drawn from the same distribution of hosts at {\color{black}a} high confidence {\color{black}(p=2.7x10$^{-5}$ and p=5.4x10$^{-5}$} for LGRBs and CCSNe respectively). We also display the \citealt[][]{Baldry2005} UV galaxy luminosity function for galaxies within our brightness range (cyan line). Using UV brightness as a proxy for SFR (top x-axis), we can expect our hosts to be substantially less star forming than our comparison samples. Breaking this down by SLSN type (lower panel) shows little distinction between the subclasses, although again small number statistics are likely to be an influence here. AD testing between subclasses proves a strong inconsistency between the all classes of SLSN hosts and our comparison samples in M$_{UV}$.}
\label{fig:cdf_uv}
\end{figure}

 Figure~\ref{fig:cdf_uv} shows the same analysis for the UV luminosity distribution of the SLSNe sample. {\color{black} Again, the SLSNe are markedly fainter (hence lower star formation rate) than the GRBs or CCSNe. However, since they are also faint in the nIR their inferred specific star formation rates (SFR/M), do not suggest that they are forming stars an a rate unusually low 
 for their mass, and they would still class as actively star forming galaxies}. Interestingly in this UV range the CCSN and LGRB hosts appear
to be more similar, although it should be noted that due to the paucity of UV observations of CCSNe in GOODS, this CCSNe host sample is different from the one used for our nIR comparison. The similarity of LGRB and CCSNe hosts in the UV, and the differences in the nIR could also be explained by the typically higher specific star formation rates of 
 GRB hosts \citep{CastroCeron2006,Svensson2010}. 
 
To formalise the significance of these differences we perform both Kolmogorov-Smirnov (KS) and Anderson-Darling (AD) tests of each population {\color{black}(including a separate tests for our 
SLSNe and combined (SLSNe+LSNe) samples)}. The AD test provides a sample comparison more sensitive to the ends of the distribution, which in light of the extremely faint nature of some of our sample, may provide a more apt test statistic than the KS test. Hence we refer to the AD statistic throughout the rest of this work, although our conclusions would be unaffected by the use of the KS-test. The probabilities of an underlying association between different distributions are presented in Table \ref{tab:AD}, and the results indicate that the probability of the SLSN host sample and the hosts of LGRBs and CCSNe being drawn from the same pool of galaxies is low. As expected the differentiation is strongest for the SLSN-I hosts, which reject the hypothesis that
they arise from hosts with similar absolute magnitudes to either CCSN or GRB
host galaxies, in both cases indicating that the host galaxies are significantly less luminous, with further implications for their masses and star formation rates (see below). 

The SLSN-II hosts have low to modest probabilities of being drawn from the same underlying host population as both the LGRBs {\color{black}($P =  0.01, 0.23$} for UV and nIR respectively) and the CCSNe {\color{black}($P = 0.008, 0.29$ for UV and nIR)}. However, as previously noted the most striking feature of the SLSN-II hosts is their presence over a wide range of luminosity from our brightest host (SN~2006gy, $M_{nIR} \sim$ -22.5) to our faintest two (SN~2008es, SN~2008fz, $M_{nIR} \sim -13$). Should these galaxies be drawn from some star formation (or mass) weighted distribution, the chance of obtaining any such faint hosts within a small sample would be very small. For example, the expected number based on the extrapolation of a luminosity function is $<<$1. Indeed, KS and AD tests suffer from a lack of sensitivity to such extremes since they measure the maximum offset between two distributions, and are insensitive to these extremes. Despite the small number statistics, the presence of two SLSNe-II in such faint host suggests than unusual mechanisms may be at play in at least some of these events.

\renewcommand{\arraystretch}{1.5}
\begin{table}
\begin{minipage}{90mm}
\centering
\caption{Properties of SLSN hosts derived from SED fitting. Uncertainties presented here are those associated with photometric errors only and do not include systematic uncertainties related to the fitted SED models. Objects marked * are detected within only one band. Mass errors provided for these objects represent for the upper and lower bound we can place upon these hosts.}
\label{table:slsn_properties}
	\begin{tabular}{l l l l l l l l l}
		\hline
		SLSN & SFR & $M_{*}$ \\
			& (M$_{\odot}$ yr $^{-1}$)	& {\color{black}($\times 10^{9}$ M$_{\odot}$)}  \\
		\hline
		\hline
		SN1995av  	&	 0.201  $^{+0.063}_{-0.077}$ &   0.578 $^{+0.270}_{-0.192}$  \\
		SN1997cy  	&	 0.170  $^{+0.207}_{-0.030}$ &   0.255 $^{+0.042}_{-0.216}$  \\
		 SN1999as  	&	 0.610  $^{+0.014}_{-0.006}$ &   2.197 $^{+0.396}_{-0.000}$  \\
 		SN1999bd  	&	 0.412  $^{+1.030}_{-0.259}$ &  10.494 $^{+1.339}_{-2.073}$  \\
 		SN2000ei	 	&	 9.597  $^{+3.511}_{-0.000}$ &   0.863 $^{+0.175}_{-0.000}$  \\
		SN2005ap  	&	 0.090  $^{+0.017}_{-0.016}$ &   0.287 $^{+0.107}_{-0.097}$  \\
 		SN2006gy  	&	 0.000  $^{+0.000}_{-0.000}$ & 153.280 $^{+6.251}_{-6.463}$  \\
 		SN2006oz  	&	 0.013  $^{+0.025}_{-0.013}$ &   0.466 $^{+0.113}_{-0.038}$  \\
 		SCP06F6*\footnote{Mass reported is an assumed fixed mass used within SED fitting}  	&	 0.136  $^{+0.028}_{-0.025}$ &   0.010 $^{+0.000}_{-0.000}$  \\
 		SN2007bi  	&	 0.048  $^{+0.006}_{-0.009}$ &   0.136 $^{+0.097}_{-0.053}$  \\
 		SN2008am  	&	 2.018  $^{+0.001}_{-0.002}$ &   5.637 $^{+0.018}_{-0.047}$  \\
 		SN2008es  	&	 0.007  $^{+0.001}_{-0.001}$ &   0.006 $^{+0.005}_{-0.005}$  \\
 		SN2008fz  	&	 0.009  $^{+0.001}_{-0.001}$ &   0.017 $^{+0.001}_{-0.001}$  \\
 		SN2009jh*  	&	 0.030  $^{+0.000}_{-0.000}$ &   0.068 $^{+0.041}_{-0.000}$  \\
 		PTF09cnd	&	 0.162  $^{+0.035}_{-0.019}$ &   0.673 $^{+0.100}_{-0.185}$  \\
 		2010gx  	&	 0.340  $^{+0.015}_{-0.018}$ &   0.349 $^{+0.055}_{-0.046}$  \\
 		 \color{black}{PTF10hgi}	&	 0.003  $^{+0.008}_{-0.003}$ &   0.351 $^{+0.020}_{-0.016} $ \\
 		SN2011kf	&	 0.174  $^{+0.061}_{-0.015}$ &   0.124 $^{+0.077}_{-0.090} $ \\
 		SN2011ke	&	 0.177  $^{+0.009}_{-0.007}$ &   0.070 $^{+0.016}_{-0.017}$  \\
 		PTF11dsf	&	 0.924  $^{+1.849}_{-0.076}$ &   2.651 $^{+0.188}_{-1.368}$  \\
 		 \color{black}{PTF11rks}	&	 0.602  $^{+0.029}_{-0.000}$ &   0.773 $^{+0.080}_{-0.000} $ \\
		SN2012il	&	 0.212  $^{+0.057}_{-0.009}$ &   0.284 $^{+0.177}_{-0.112}$ \\

		\hline
 	\end{tabular}
\end{minipage}
\end{table}
\renewcommand{\arraystretch}{1.0}
 
\begin{table*}
\begin{minipage}{120mm}
\centering
\caption{Two sample Anderson-Darling probability results between samples. Probabilities $\textless\sim\times$10$^{-6}$ are given 0.0}
\label{tab:AD}
	\color{black}
	\begin{tabular}{l l l l l l}
		 \hline
		 
		 & & \multicolumn{2}{c}{SLSNe Sample} &  \multicolumn{2}{c}{Combined Sample\footnote{Excluding SN~1997cy}} \\
		\hline
		Data Set & Host Connection & KS & AD & KS & AD \\
		 & Connection & Stat. & Stat. &Stat. & Stat.\\
		\hline
		\hline
						& SLSNe - LGRB 	&		0.013 & 0.008 &	0.022 &  0.020	\\
		nIR Magnitude 	        & SLSNe - CCSNe 	&		0.005 &  0.0017  &	0.009	& 0.003 \\
						& SLSNe-I - LGRB 	&		8.1x10$^{-4}$ &  8.1x10$^{-5}$	& 1.5x$10^{-4}$& 8.7x10$^{-5}$\\
						& SLSNe-I - CCSNe   & 		1.5x10$^{-5}$ & 6.8x10$^{-5}$	&4.4x10$^{-5}$ &  6.8x10$^{-5}$	\\
						& SLSNe-II - LGRB 	&		0.55 & 0.23 & 0.77 & 0.33 	\\
						& SLSNe-II - CCSNe  &		0.61 &  0.29				&0.56  &  0.27\\
						& LGRB - CCSNe	&		0.05 &  0.04			& - & - \\

		\hline
						& SLSNe - LGRBs 			&  4.5x10$^{-5}$& 	2.7x10$^{-5}$		 & 3.2x10$^{-5}$		&  3.1x10$^{-5}$			\\
		UV Magnitude 	       & SLSNe - CCSNe 			& 1.4x10$^{-5}$ & 	5.4x10$^{-5}$		& 1.0x10$^{-5}$& 6.5x10$^{-5}$	\\ 
						& SLSNe-I - LGRB 			& 1.0x10$^{-6}$& 1.9x10$^{-6}$    	& 1.0x10$^{-6}$ &1.4x10$^{-5}$	\\
						& SLSNe-I - CCSNe 			& 1.7x10$^{-6}$&   5.0x10$^{-5}$   		& 0.0 &2.7x10$^{-5}$		\\
						& SLSNe-II - LGRB 			& 0.12 & 0.01 & 0.07 & 0.009	\\
						& SLSNe-II - CCSNe  		& 0.06 & 0.008 & 0.053 &  0.011	\\
						& LGRB - CCSNe			& 0.85 &	0.84		& - & -\\
		\hline
						& SLSNe - LGRB 		 	& 0.0 & 0.09 & 0.0 & 0.14	\\
		Masses			& SLSNe - CCSNe 		 	& 0.0 & 0.015 & 0.0  & 0.0019	\\
						& SLSNe-I - LGRB 		 	& 0.002 & 1.2x10$^{-5}$& 8.6x10$^{-4}$ & 0.0	\\
						& SLSNe-I - CCSNe 		 	& 8.9x10$^{-5}$ & 0.0& 2.6x10$^{-5}$& 0.0 	\\
						& SLSNe-II - LGRB 			& 0.64 & 3.3x10$^{-5}$& 0.84 & 1.5x10$^{-5}$	\\
						& SLSNe-II - CCSNe  	  	& 0.49& 1.32x10$^{-5}$& 0.70 & 0.0	\\
						& LGRB - CCSNe			& 0.48 &	 0.12		& - & - 	\\
		\hline
						& SLSNe-LGRB 			 & 6.7x10$^{-5}$&6.9x10$^{-5}$ & 5.5x10$^{-5}$ &	9.9x10$^{-5}$\\
		SFRs 			& SLSNe - CCSNe 			 & 0.0 & 0.0& 0.0 & 0.0 	\\
						& SLSNe-I - LGRB 			 & 8.3x10$^{-5}$&2.6x10$^{-4}$ & 9.1x10$^{-5}$&1.5x10$^{-5}$	\\
						& SLSNe-I - CCSNe 			 & 0.0 & 1.34x10$^{-5}$& 0.0 & 1.1x10$^{-5}$	\\
						& SLSNe-II - LGRB 			 & 0.065 & 0.065 & 0.09 &	 0.02\\
						& SLSNe-II - CCSNe  		 & 0.016 & 0.0013 & 0.04 & 0.002	\\
						& LGRB - CCSNe			& 0.11 &	0.06		& - &	-\\
		\hline
		r$_{80}$ 			& SLSNe - LGRB 			& 0.0 & 1.0x10$^{-5}$ & 	0.0 & 1.5x10$^{-5}$		\\
						& SLSNe - CCSNe 			& 1.4x10$^{-4}$			 & 	 1.1x10$^{-4}$		& 7.3x10$^{-4}$ & 3.6x10$^{-4}$	\\
						& LGRB - CCSNe			& 0.15  &	 0.09			& - & - \\
		\hline
 	\end{tabular}
\end{minipage}
\end{table*}

\begin{figure*}
\begin{minipage}{180mm}
	\centering
	\includegraphics[scale=0.9,angle=0]{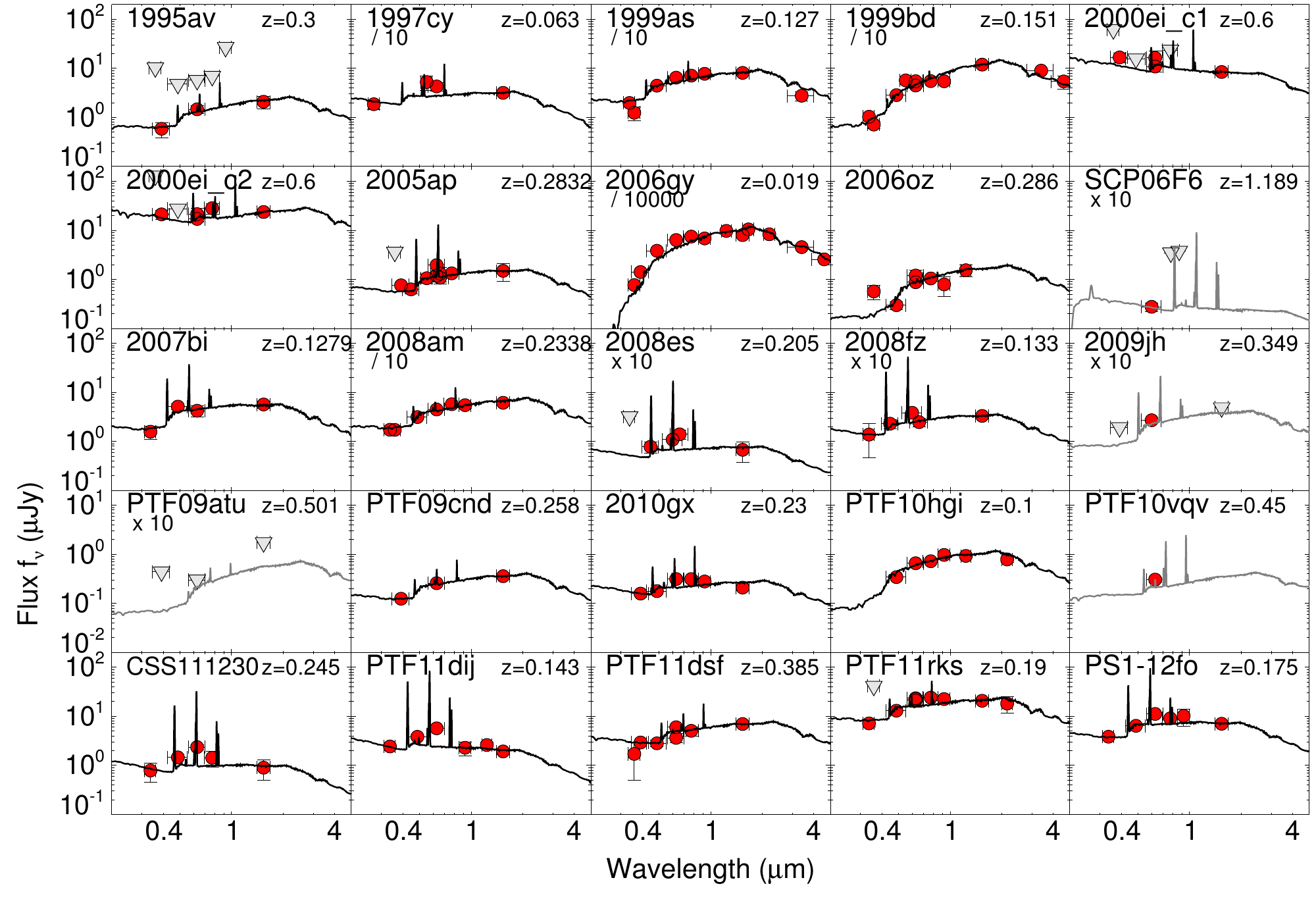}
	\caption{SED fits of SLSN hosts as carried out in a similar manner to \citealt[][]{Perley2013}, using photometric results from our HST programmes and WHT and VLT images, in addition to results from literature and SDSS. Arrows indicate upper limits to photometry}
	\label{fig:SED}
\end{minipage}
\end{figure*}

\begin{figure}
	\centering
		\includegraphics[scale=0.45,angle=0]{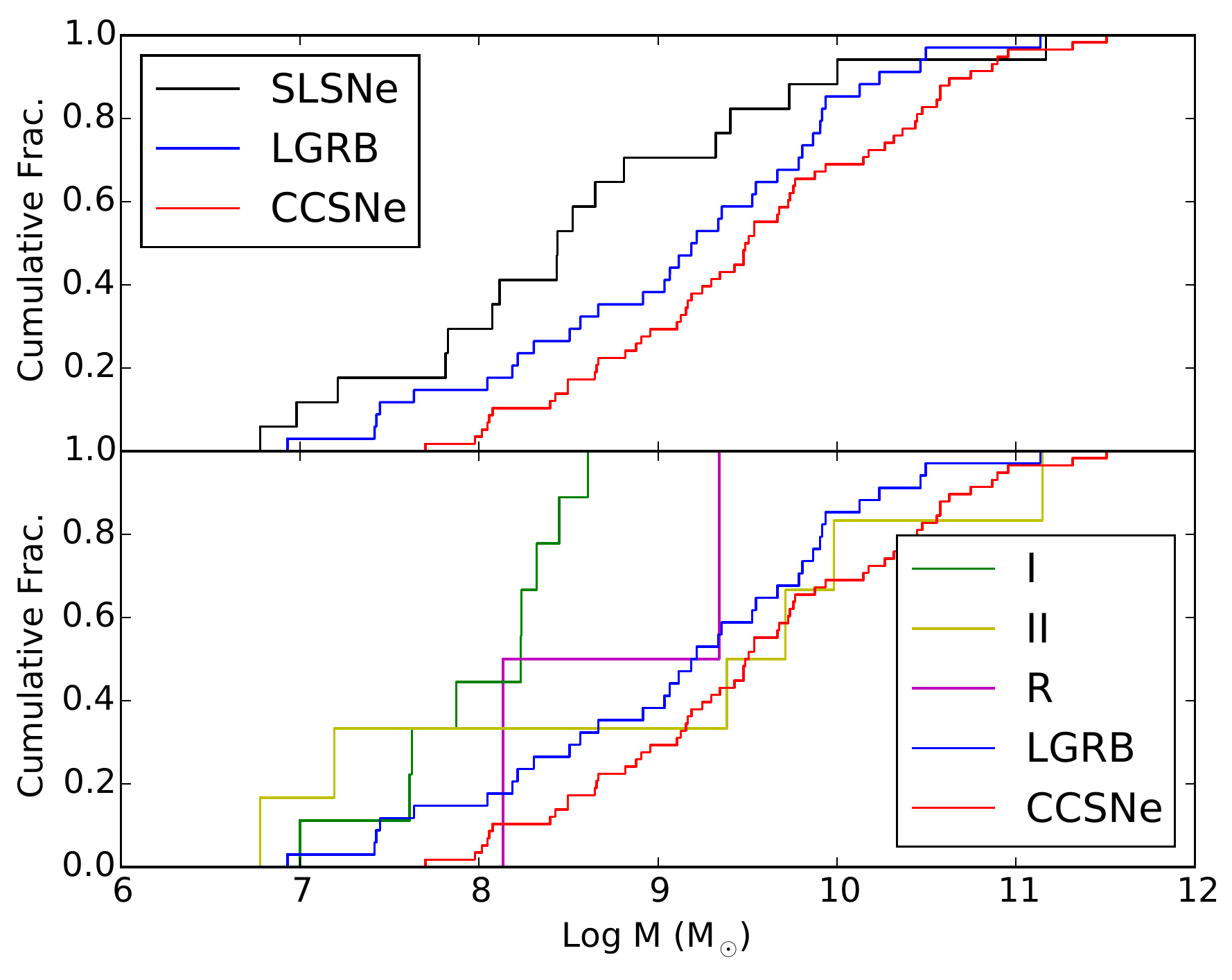}
	\caption{Masses of transient hosts as determined by SED fits. SLSN hosts are significantly less massive than CCSNe host galaxies, and show a 1$\sigma$ difference to LGRB hosts. Splitting by subtype we again find little similarity between the subclasses of SLSN hosts and our comparison samples}
	\label{fig:cdf_mass}
\end{figure}

\begin{figure}
	\centering
		\includegraphics[scale=0.45,angle=0]{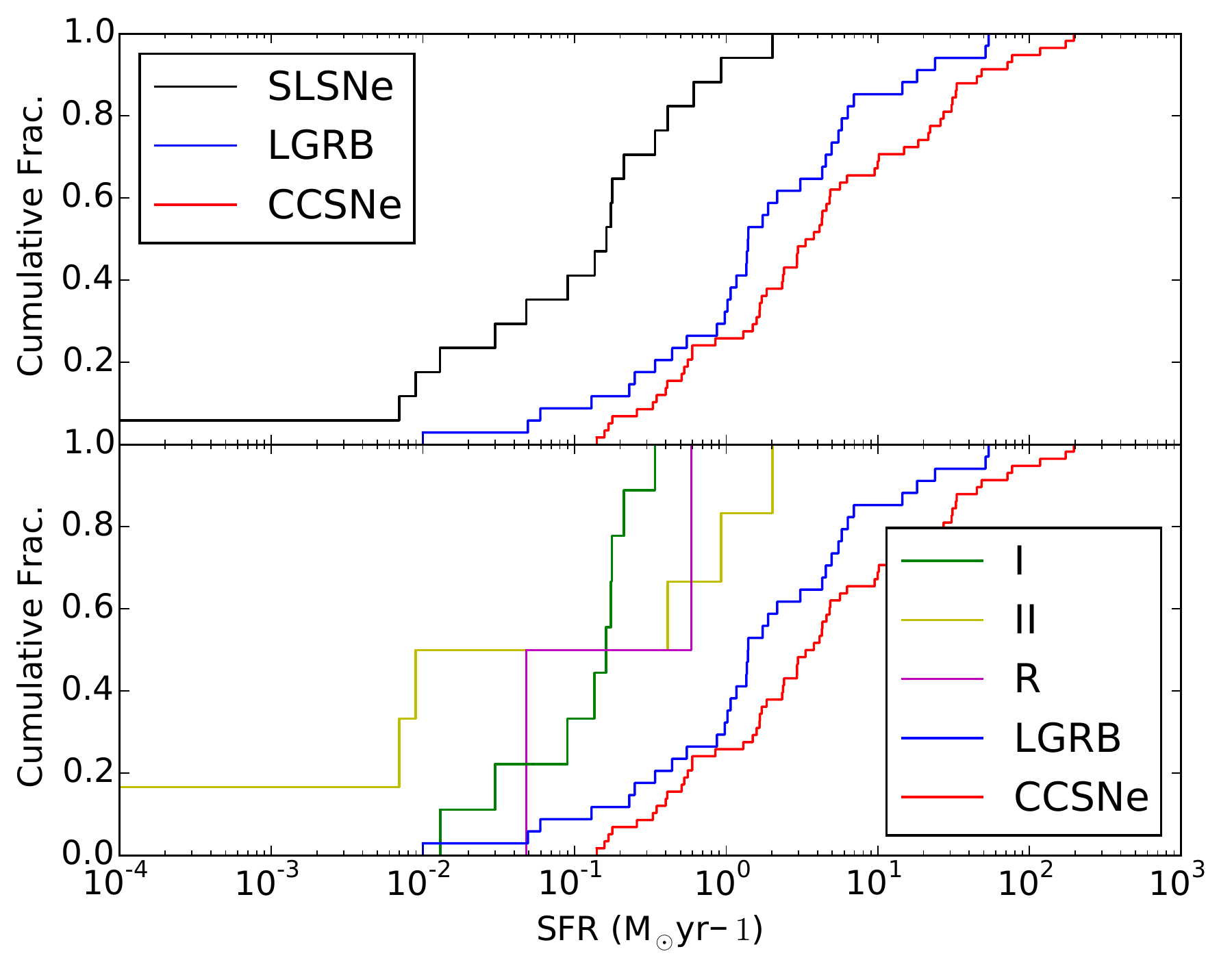}
	\caption{Star formation rate for transient hosts determined by SED fitting. SLSN hosts are seen to be not as strongly star forming as CCSNe or LGRB host galaxies, with very low probabilities of the distributions being from the same underlying population. Again, splitting by subtype shows little deviation from this result for the hosts of SLSNe-I, however for those of SLSNe-II, there appears to be slight overlap between it and the LGRB hosts distribution (p=0.08). However, the evolution exhibited within the average SFR of LGRB hosts over low redshift \citealt[][]{Perley2013,Perley2015b,Schulze2015}, may somewhat bias our results.}
	\label{fig:cdf_SFR}                                                          
\end{figure}

We present our SED fits {\color{black}to all our targets} in Figure \ref{fig:SED}, and our
derived properties in Table~\ref{table:slsn_properties}. We compare these stellar masses and 
star formation rates to those found through proxies from our nIR and UV luminosities, which provides a model independent check upon our SED fit values, and find them to be generally of the same order of magnitude. Using the properties derived from the SED fitting, we present the distribution of masses and SFRs for our sample in Figures \ref{fig:cdf_mass} and \ref{fig:cdf_SFR} respectively, alongside those properties which have been derived from SED fitting for LGRB and CCSN hosts from \cite{Fruchter2006} and \cite{Svensson2010}.  As suggested by proxies, we can see that SLSN hosts  are less massive and possess lower SFRs than CCSN and LGRB hosts, to a high level of significance, as show in Table \ref{tab:AD}. 

\begin{figure*}
\begin{minipage}{120mm}                   
	\centering
		\includegraphics[scale=0.6,angle=0]{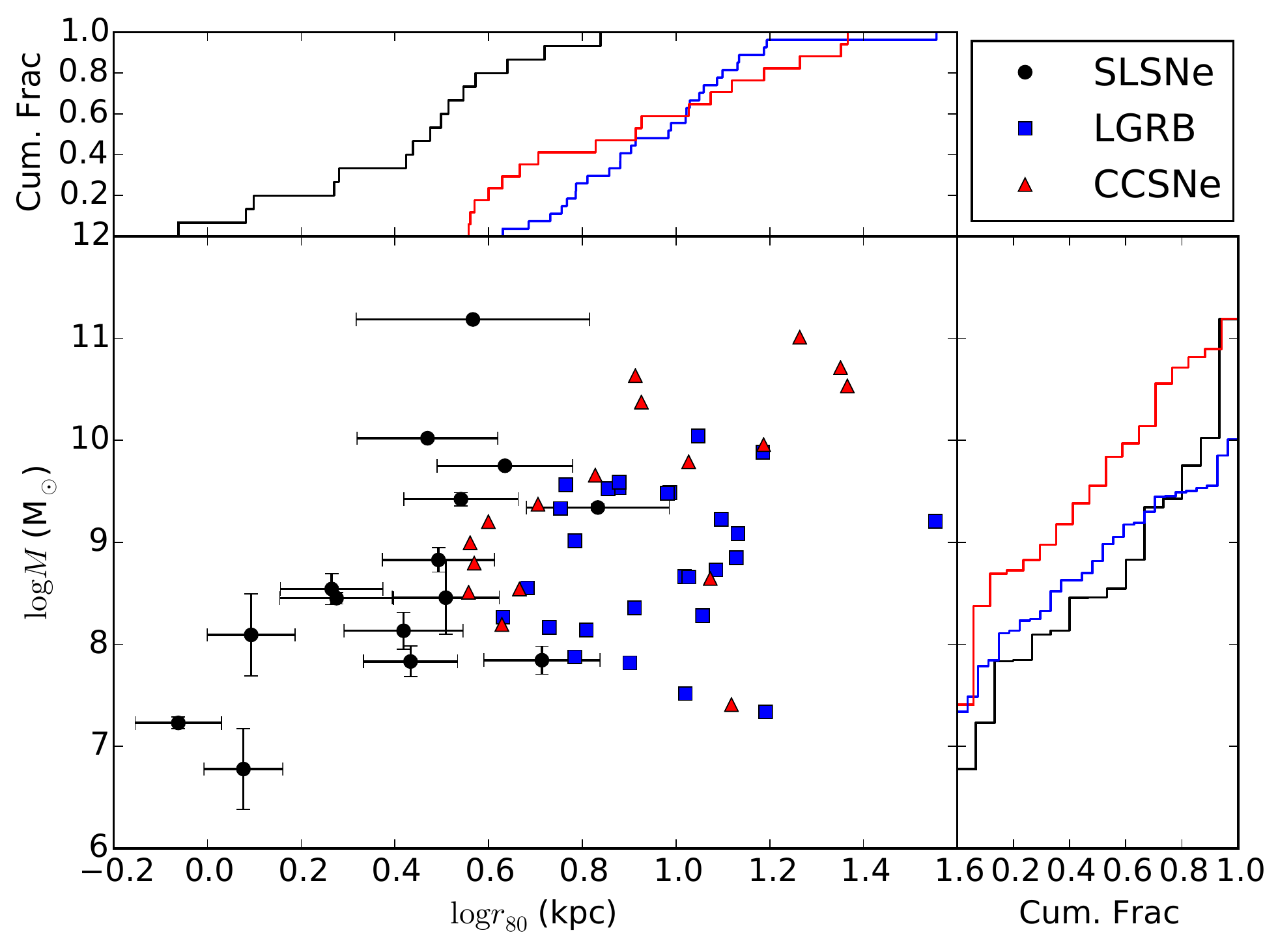}
	\caption{r$_{80}$ light profiles measured in the HST F160W band of core collapse hosts against their mass as derived from SED fitting. Error bars are indicative of SExtractors ability to detect the edge of a galaxy at given brightness for a given redshift. The compact nature of SLSN hosts is apparent here.}
	\label{fig:r80}
\end{minipage}
\end{figure*}

A comparison of the measured r$_{80}$ values from our nIR observations is presented in Figure \ref{fig:r80}, combined with the masses to provide an indication of the relative evolution of size with luminosity for our core collapse transient host sample. The compact and low mass nature of the SLSN hosts is clearly visible, as they occupy a distinct region of parameter space from other core collapse hosts of similar brightness. We note that CCSN hosts are in turn more compact than SDSS galaxies \citep[e.g.][]{Kelly2010}, whose size distribution peaks well above of the range of sizes presented within this work.  Again, AD tests between the HST SLSNe and comparison samples give little probability that they are drawn from the same
underlying population. It should be noted that our nIR observations are frequently rather short, 
and so low surface brightness features could be missed in comparison to deeper observations of the
GRB hosts and GOODS SNe. However, we evaluate the probability of this using modelling of galaxies to estimate our expected recovery rate. We find that even if SLSN hosts were to lie at the extrema of their error bars (i.e. 
if there were a systematic shift of each point by 1$\sigma$ larger) the result would still be statistically
significant to 1x10$^{-4}$ and 0.014 for LGRBs and CCSNe respectively. 

Given that the redshift distributions of these classes of transient exhibit somewhat different functional forms it is reasonable to ask if the observed differences in the properties of the population are due to redshift evolution in the host galaxies, rather than the properties of the progenitor stars themselves. 

Ideally it may be beneficial to conduct tests considering only low-z SLSNe (e.g. $z<0.4$) and with comparison samples at the same redshift. However, our comparison samples become very small at these low-redshifts, frequently with $<4-8$ objects for comparison ({\color{black} see Figure 1}). For these small sample sizes we lack the statistical power to make strong statements about redshift evolution within the SLSNe sample in comparison to those of others. Given that there is some evidence for evolution in LGRB properties with redshift, albeit occurring predominantly around $z\sim 1$ \citep{Perley2015b} it is possible that some apparent differences between SLSNe and other transient populations are amplified, or damped, by evolution in the host properties themselves.

\begin{figure}
	\centering
		\includegraphics[scale=0.45,angle=0]{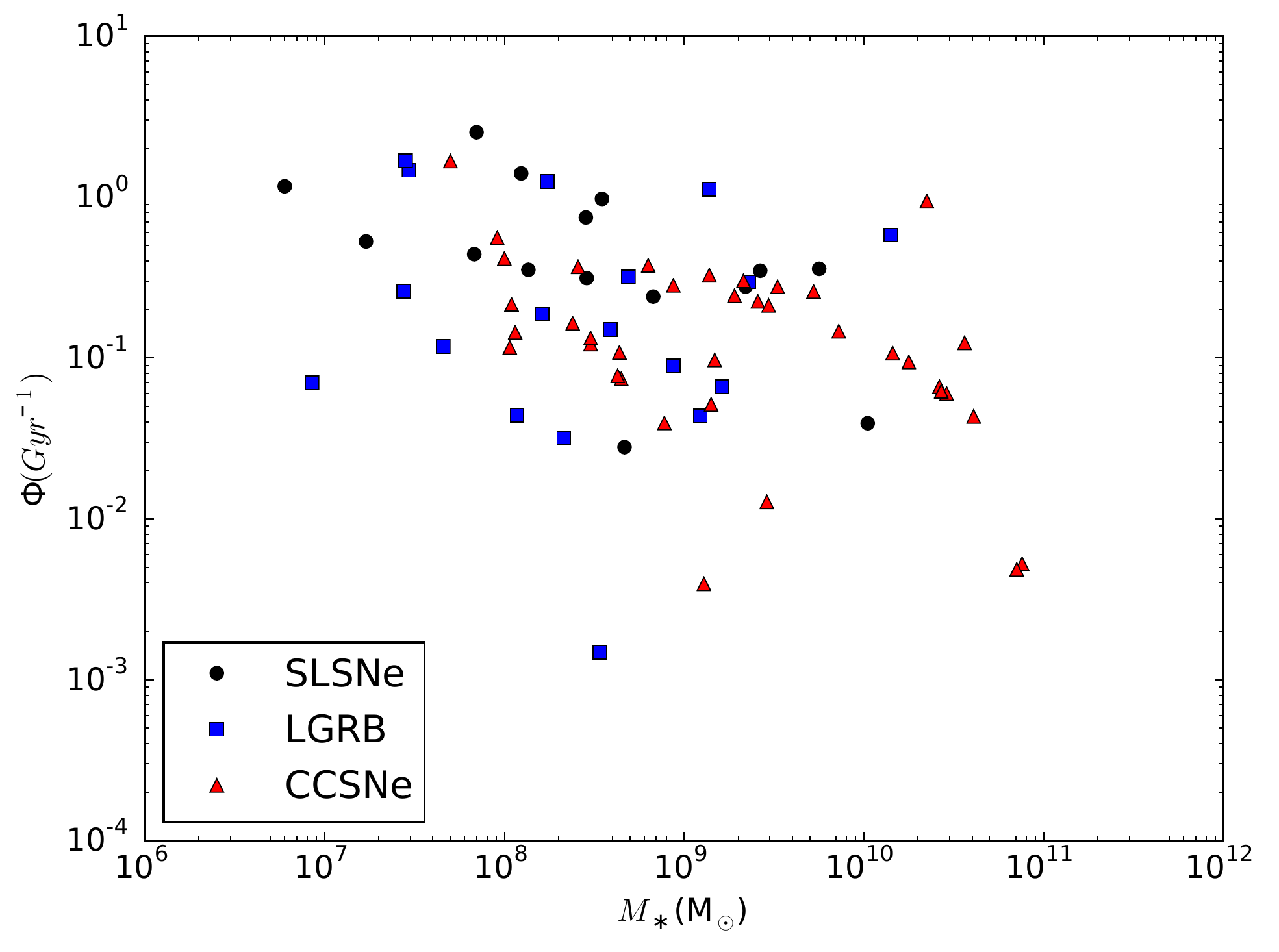}
	\caption{sSFR values for SLSN, CCSN and LGRB hosts against their respective stellar masses. Overall, the hosts of SLSNe appear to occupy a similar range of sSFR values as CCSN and LGRB host galaxies. \color{black}{Note we do not include the host of SN~2006gy here, due to it's poorly constrained star formation rate from SED fitting.}}
	\label{sSFR}
\end{figure}

We also determine  specific SFRs (sSFRs) for our SLSN hosts, which we present within Figure \ref{sSFR}. When compared alongside those of LGRBs and CCSNe from \cite{Svensson2010}, appear to fall within a similar range of sSFR as other core collapse transients. Although, when compared to a wider sample of galaxies, as carried out by \cite{CastroCeron2006} (ref. their figure 2) and \cite{Svensson2010} (ref. their figure 7), such as distant red galaxies (DRGs), submillimeter galaxies (SMGs) and Lyman break galaxies (LBGs), the sSFRs of the core collapse transients lie at lower masses for a given sSFR than DRGs, SMGs and LBGs. 

\section{Discussion}

The observations presented within this work highlight the extreme nature of the SLSN host population.  A significant fraction arise in galaxies of exceptionally low luminosity, both in the UV and nIR. {\color{black} These galaxies are extreme even when compared to other populations of core collapse hosts, or even to GRBs, whose host galaxies are already set well apart from a typical field sample.} Given that the UV and IR naturally provide a probe of both star formation and stellar mass, these differences are indicative of extremely low mass star forming hosts for SLSNe. Indeed, studies of SDSS galaxies indicate that there is little contribution to the global star formation rate in the local Universe from galaxies with M$_{UV} > -17$ \citep{Blanton2005,Graham2013}, where we have shown the majority of the SLSN hosts within our sample lie, as can be seen in Figure~\ref{fig:cdf_uv}. This result also holds in comparison to the host galaxies of CCSNe and LGRBs, the latter of which
have been suggested to arise predominantly, if not exclusively from stars of low to moderate metallicity
\citep[e.g.][]{Fruchter2006,Graham2013,Perley2015a}. 
The host galaxies are also typically small, but
exhibit surface star formation densities, and specific star formation rates that are more in keeping with
those of other transient populations (i.e. they lie at the low end of most physical parameters compared to other core collapse transient hosts, such that
any additional parameter derived with reference to two of more of SFR, mass and size, does not provide a strong distinction between the hosts of SLSNe and other star forming galaxies). The majority of our hosts exhibit high star formation surface densities, higher than those seen in the hosts of SNe-Ib/c and SNe-II, more akin to broad line SN-Ic and GRB hosts \citep{Kelly2014}, in agreement with the results of \cite{Lunnan2015}.

However, these broad conclusions based on all SLSNe fail to consider the
 diversity of SLSN types. In splitting the sample by type (utilising the classification system
 of \citealt[][]{Gal-Yam2012}),
small number statistics prevent us from drawing
strong conclusions about differences {\em between} SLSN subtypes, although it does 
appear that SLSNe-I arise from predominantly fainter host galaxies than SLSNe-II on average. 
The larger differences between SLSNe and other classes of transient (compared to the
differences between classes of SLSNe), do allow us to draw stronger conclusions when comparing
the host galaxies of 
SLSNe-I and SLSNe-II to the hosts of LGRBs and CCSNe.

The SLSN-I hosts are much fainter than the hosts of either CCSNe or LGRBs. Since the
LGRBs are frequently explained as arising from low metallicity systems, the logical conclusion
might be to assign SLSNe-I to progenitors of even lower metallicity. This however is problematic;
spectroscopic observations of the hosts of SLSNe \citep{Lunnan2014,Leloudas2015} 
generally show modest metallicities, and indeed \cite{Lunnan2014} conclude the metallicities of SLSNe-I are consistent with those of GRB hosts. There are multiple
possible origins for this discrepancy. 

Firstly, it may be that rapid evolution in the properties of LGRBs hosts with redshift magnifies what is in fact a small difference between the metallicity cuts for SLSNe and LGRBs. Although small sample sizes prevent us from testing this reliably, it is not unlikely that evolution within the LGRB host population below $z\sim 1$ may accentuate the apparent differences between themselves and SLSN hosts. Additionally, the samples utilised by \cite{Lunnan2014}, \cite{Leloudas2015} and this work, while containing some overlap are also significantly different. Small number
statistics may then represent a potential concern. 

Selection effects could also hinder such work. For example, many SLSNe have been found by searches targeting orphan transients (those without visible hosts in the survey images), since the SLSNe so effectively outshines it host galaxy. This may
immediately remove SLSNe in higher metallicity, more luminous hosts, causing the 
remaining sample to be biased towards a lower metallicity.  We can attempt to address this by adopting the PanSTARRS limiting magnitude cut of R$\sim$23.5 for host galaxy detection across all of our host samples  (SLSNe, CCSNe and GRBs), such that we include {\textit{only}} hosts fainter than this limit  (we note that this is the most conservative approach since the limiting magnitudes of the other surveys finding SLSNe are typically significantly brighter). We recover 8/21 hosts from our HST SLSN sample  using this approach. Within this limit the SLSN host sample appears fainter and less massive than the CCSNe and LGRB host samples. Although here we are once again dominated by small number statistics within our comparison hosts it suggests that the differences between the differing populations are not created by the selection mechanisms of the transient surveys. The impact of the faintest 
galaxies may operate in the opposite direction, very faint galaxies are difficult to obtain 
metallicities for, and so if these are omitted it may bias the observed metallicity distribution
towards higher levels. 

Finally, it is relevant to consider if astrophysical effects could be at play. {\color{black} 
Mass (or luminosity) metallicity relations have been used to infer the metallicities of GRB host galaxies, and
this could be extended to SLSN hosts. In this case one might infer a metallicity threshold based on the
most luminous observed SLSN host galaxy, and could then test the consistency of the distribution of
fainter (and using an L-Z relation, lower metallicity) galaxies. In this case the observed distribution 
of SLSNe-I would be broadly in keeping with expectations. For the UV luminosity function of \cite{Baldry2005},
truncated at $M_{UV} \sim -16.8$ (our most luminous SLSN-I host)
we would expect $\sim 60\%$ of the UV-light (hence SFR, or equivalent number of SNe) to arise from
galaxies within 1-magnitude of this luminosity. This would match well the relatively narrow range of 
luminosities observed for the host galaxies of SLSNe-I, while the 2 upper limits (of 9 SNe) are consistent
with the fainter fraction of the hosts. To this end, metallicity may appear an good description of the 
observed luminosity distribution. However, 
it is clear such relations between luminosity or mass and metallicity are} crude at best;  often GRB hosts are found to have low metallicity, even when in
relatively luminous hosts (see e.g. figure 10 in \cite{Graham2013}). If SLSN hosts lie systematically low in metallicity when compared to mass in the mass metallicity relationship then it would not be surprising that they could appear
very different from LGRBs in mass, but rather more similar in metallicity. It is also possible that an apparent discrepancy in interpretation may arise due to the different locations of SLSNe and GRBs on their host galaxy light distributions. GRBs are preferentially concentrated
on the brightest regions of their host galaxies. In these situations the global metallicity of the host
galaxy (which comes from ``most" of the light) might be a reasonable proxy for the metallicity in the GRB region (although see e.g. \cite{Hammer2006} for some caveats). In the case of SLSNe, the 
concentration is not so strong \citep{Lunnan2015}, and indeed some events (e.g. SN2009jh) lie
apparently off their host galaxy light. In these scenarios it is more likely that the global host
metallicity is not indicative of the metallicity at the location of the SNe, and so spatially resolved
measurements are urgently needed. 

Theoretically, there are good reasons to favour similarities between the environments of 
LGRBs and SLSNe-I. It is known that LGRBs arise from central engines \citep{Woosley1993}, and there is growing consensus that this is also the case for SLSNe-I \citep{{Kasen2010},{Dexter2013}}, which become active during the collapse of very massive stars. Observations of both classes of event provide evidence favouring this model (for SLSNe \citep{{Levan2013},{Nicholl2013},{Nicholl2015A}}, for LGRBs e.g. \citealt[][]{Metzger2011}, although for association with luminous SNe, see \citealt[][]{Mazzali2014}). If this is the case
then we might expect the production of these engines to be favoured in similar environments. However, 
there are differences in the necessary engine properties to create LGRBs or SLSNe. In particular, in LGRBs, the bulk of the energy must be released extremely early ($\sim 10 ^{3}$s) to power the ultra-relativistic outflow, this energy is then deposited into the ejecta close to the engine. In contrast, for SLSNe the engine must act to re-engerize the outflow on timescales of weeks to months after the initial core collapse. In the case of black hole engines this means the accretion timescales must vary by many orders of magnitude, while for magnetars the crucial spin down parameter must also be different. 

 Relative numbers of magnetars observed within the Milky Way, when placed in context with the galactic CCSN rate, requires that $\sim$10\% of these events result in the birth of a magnetar \citep{Mereghetti2015}.  This rate is far higher than any suggested for SLSNe and suggests that the magnetars we 
observe in the Galaxy today have little connection to those that may be created in luminous SNe explosions. Rotation is a logical difference between those systems creating ``normal" magnetars, and those which are powerful enough to re-energize explosions, and this may in turn provide an natural explanation for environmental biases. At higher metallicities the line driven winds
will dramatically brake the rotation of the star prior to a supernova explosion, and hence conservation of magnetic flux and angular momentum upon core collapse may create a 
magnetar with a longer rotation period than needed to explain either GRBs or SLSNe.  Hence we might expect to observe both LGRBs and SLSNe in relatively
metal poor environments. Indeed, since
the spin periods for the GRB magnetars are shorter than for those creating SLSNe (or they have higher magnetic fields) one might naively assume that GRBs could favour even lower metallicity. In this regard it is valuable to note the recent example of GRB 111209A, an ultra-long GRB in a low metallicity galaxy \citep{Levan2014} in which a magnetar may have produced both the GRB and luminous SNe \citep{Greiner2015}.

We have only two SLSNe-R within our sample, and so can say little about the properties of their
hosts in comparison to other samples, aside from noting that their luminosity is generally
in keeping with those of SLSNe-I, which some authors have suggested is their correct
assignment. We do note that interestingly in both cases the SLSNe-R appear to originate
from bright UV regions within their hosts, something that is not the case for all SLSNe-I, but given
the small sample size and available data it is not possible to investigate if they may arise from young, massive stellar populations in metal poor regions, more so than the environments of SLSNe-I. It is also relevant to note that
recent calculations suggest that stars at modest metallicity and mass can create
pair instability SNe \citep{Yusof2013} and so the environment alone may not ultimately
provide as strong a means of discrimination between models as had previously been hoped.

Less attention has been paid to the host galaxies of SLSNe-II, partly as the interaction model for their
origin appears a more natural explanation given the likely presence of recently ejected hydrogen envelopes in Type-IIn SNe (and most SLSNe-II are of the IIn variety). However, their hosts span a very wide range of luminosity, including two host galaxies that are fainter than any SLSNe-I, LGRB or CCSN host in our sample. Indeed, while a handful of SNe Ia have been found
in comparably faint systems \citep[e.g.][]{Strolger2002} the presence of any
type of core collapse SNe 
in galaxies fainter than M$_B \sim -14$ is extremely rare (for example,
none in the cross correlation of the SAI catalog with SDSS \citep{Prieto2008}). Although
this may in part be due to a lack of follow-up, in practice at these
modest redshifts essentially no SNe would be expected, even with the metallicity cuts used
to explain the GRB population \citep{Graham2013}. The presence of two host galaxies in such low 
luminosity galaxies is then puzzling; whatever mechanism is at play must be able to produce supernovae across this wide range of galaxy types. Metallicity dependence here seems a less likely scenario, unless those
SLSNe-II apparently born in the most luminous hosts are in fact born in lower mass dwarf galaxies within their halos (although in this case it would be odd that some  SLSNe-I were not also seen in
similar environments). However, other possible mechanisms may provide a viable alternative. For example, if SLSNe-II were formed only from very massive stars then they may exist only in very special locations. If the core mass prior to supernova is the dominant factor then indeed low metallicity will preserve core masses much better than
at higher metallicity due to far lower radiative mass loss rates, and a possible bias
to a more top heavy initial mass function at lower metallicity. However, if SLSNe-II are in fact best explained by a strong interaction model then
large scale mass loss is necessary at some point. In this case, the conditions necessary to form a SLSNe-II may be a combination of 
both relatively high core mass and still significant mass loss, meaning the initial (i.e. total) mass could play a more important role. In this regard it is interesting to note
that the formation of very massive stars is potentially affected by stochastic processes even without changes in metallicity or to the underlying IMF. 
Small star forming regions, following a typical initial mass function, {\color{black}have a lower probability of building} most massive stars, because there is insufficient mass. For example, if a star forming region will form 
only a few hundred solar masses of stars the probability of it forming any stars with greater than $\sim 
100$ M$_{\odot}$ is extremely small, {\color{black} stochastic sampling assumes that masses are picked at random from the IMF, but that the star can only be formed should sufficient mass remain in the cluster. Hence, once a few stars have been formed, forming extremely massive stars in low mass clusters becomes unlikely. Stochastic sampling effects have been 
observed in relatively local open clusters, and appear to be very important below cluster masses of $\sim 10^4$ M$_{\odot}$ \citep{Piskunov2009}}. Indeed, the most massive star in a cluster is thought to scale roughly as $0.39 M{_{\text{cluster}}}^{2/3}$ \citep{Bonnell2001,Kroupa2011}, meaning that clusters with initial masses of $\sim 10^4$ M$_{\odot}$ are needed to form stars with masses $>200$M$_{\odot}$. The most massive stars would then be formed in locations where either there was a large scale starburst (e.g. the very massive stars located in 30 Dor, or at a
handful of locations within the Milky Way \citep{{Rauw2004},{DeBecker2006},{Crowther2010},{Gvaramadze2013},{Hainich2014}}), or in places where the IMF was
biased towards the creation of high mass stars (i.e. was top heavy relative to the local IMF). Indeed, it is interesting to note that the relative number of high mass clusters (scaled by star formation rate) does appear to be higher in dwarf galaxies, or in starbursts \citep[e.g.][]{Bastian2008}, such that massive clusters, and hence the most massive stars may be found in relatively greater numbers in these galaxies, compared to relatively quiescent spirals such as the Milky Way.  Qualitatively this model may have some appeal in explaining the unexpectedly large range of properties in the SLSN-II host population, although the lack of knowledge about variations in the IMF, even in the
relatively local universe precludes more detailed work. Finally, it is also possible that multiple progenitor routes are at play in the creation of the SLSNe-II population, meaning that some exhibit strong metallicity biases while others are formed at more typical metallicities, perhaps via binary interactions which may
eject large mass reservoirs quickly during common envelopes etc.

\section{Summary}
We have utilised the unparalleled UV and nIR sensitivity of {\em HST} to provide
rest-frame UV and nIR observations of a sample of SLSNe. We find that the hosts of SLSNe-I are 
consistently fainter than other core collapse hosts (CCSNe and LGRBs), by extension this should be
indicative of a
low mass, star formation rate and metallicity. This is despite apparently similar metallicities observed
between LGRBs and SLSNe-I from optical spectroscopy of SLSN hosts (including some
hosts for which nIR and UV observations are presented here, \citealt[][]{Lunnan2014}). This discrepancy may be explained by 
a combination of small sample sizes and the absence of the faintest host galaxies from spectroscopic samples, although despite the similarities in the favoured progenitors for LGRBs and SLSNe-I there
are also good astrophysical motivations (for example the timescales required in energy breakout and potentially the spin-down rate of any magnetar driven engines) as to why their environments may not be identical. 

We find that SLSNe-II arise from galaxies spanning a surprisingly large range in absolute magnitude (and hence in star formation rate and stellar mass). This is difficult to explain from sampling the underlying star forming galaxy population subject to a simple metallicity bias, as has been attempted for LGRBs and SLSNe-I, but may be due to the preferential production of very massive stars in certain environments (either massive star formation regions, or at low metallicity). Equally, it could be a reflection that the current classification system has failed to adequately capture the true diversity of progenitor routes for SLSNe-II. 

Nevertheless it is clear that studies of SLSNe environments may still offer a powerful route to clues to their progenitor characteristics, in much the same way as they have for other classes of astrophysical transients. Such work will rely on a continuing stream of these very rare events, coupled with detailed follow-up across the electromagnetic spectrum. Through this detailed study of the
environments we may hope to elucidate the progenitors of SLSNe, and how they fit in to the growing 
diversity now being discovered in the transient optical sky.

\section*{Acknowledgments}
CRA acknowledges receipt of a studentship from the Midlands Physics Alliance. AJL thanks STFC for support
under grant ID ST/I001719/1, and the Leverhulme
Trust for support via a Philip Leverhulme Prize. We are grateful to the PTF group (in particular Robert Quimby and Avishay Gal-Yam) for providing 6 images for SNe for astrometric purposes. ALJ, JDL and ERS are grateful for the support of STFC Warwick Astrophysics consolidated grant ID ST/L000733/1.
Based on observations made with the NASA/ESA Hubble Space Telescope, obtained at the Space Telescope Science Institute, which is operated by the Association of Universities for Research in Astronomy, Inc., under NASA contract NAS 5-26555. These observations are associated with program GO-13025 and GO-13480. Based on observations
made with ESO Telescopes at the La Silla Paranal Observatory
under programme ID 092.D-0815. Optical data were obtained from the William Herschel
Telescope under service mode programme SW2012b31.

\bibliographystyle{mnras}
\bibliography{References} 
\appendix

\section{Appendix A} \label{AppA}

\begin{table*}
\begin{minipage}{120mm}
\centering
\caption{SLSNe discovery images used for carrying out astrometry and identifying host galaxies. Details for images taken from literature can be found in the following sources $^{1}$ \citet{Rawlings1996}, $^{2}$ ESO 59.A-9004(A), Service Mode, NTT , $^{3}$ Rezman Observatory ,{\color{black}$^{4}$ GO-10877, PI: Li },$^{5}$ \citet{Barbary2009},$^{6}$ \citet{Gal-Yam2009} ,$^{7}$ \citet{Chatzopoulos2011} ,$^{8}$ ToO ESO ,$^{9}$ Service Mode, NTT ,$^{10}$ Discovery images courtesy of PTF  ,$^{11}$ ToO Gemini South, $^{12}$ ToO Gemini North  }
\label{tab:discovery}
	\begin{tabular}{l l l}
		\hline
		SLSN 			& Ref. Image 	& Instrument    \\
			  			& 		  		\\
		\hline           
		\hline         
		SN1995av   		& $^{1}$		& WHT     \\
		SN1997cy   		& -				&   -       \\
		SN1999as   		& $^{2}$		& ESO NTT       \\
		SN1999bd   		& -				&   -       	\\
		SN2000ei 	 		& -				&   -       \\
		SN2005ap   		& $^{3}$		& RezmanI      \\
		SN2006gy   		& $^{4}$		& {\color{black}HST} \\
		SN2006oz    		& -				&   -       	\\
		SCP06F6 	& $^{5}$		& HST      		\\
		SN2007bi    		& $^{6}$		& ESO FORS2 VLT \\
		SN2008am    		& $^{7}$		& ROTSE Keck   \\
		SN2008es    		& $^{8}$		&      			\\
		SN2008fz    		& $^{9}$		& ESO NTT     \\
		SN2009jh    		& $^{10}$		& PTF/P60        	\\
		PTF09atu  		& $^{10}$		& PTF/P60        \\
		PTF09cnd  		& $^{10}$		& PTF/P60        \\
		SN2010gx    		& $^{11}$		& GMOS Gemini-S \\
		 \color{black}{PTF10hgi}  		& -				& -        	\\
		PTF10vqv  		& -				& -        	\\
		CSS111230 		& -				& -        	\\
		PTF11dij  		& $^{10}$		& PTF/P60        	\\
		PTF11dsf  		& $^{10}$		& PTF/P60          	\\
		 \color{black}{PTF11rks}  		& $^{10}$		& PTF/P60        	\\
		SN2012il  		& $^{12}$		& GMOS Gemini-N \\

		\hline
 	\end{tabular}
\end{minipage}
\end{table*}

\begin{table*}
\centering
\caption{CCSNe comparison sample selected from the GOODs survey, for which we carry out photometric measurements, with redshifts and positions included. }
\label{table:IR_CCSNe_Comp}
	\begin{tabular}{lllllllll}
	 	\hline
		Event & Redshift & RA(J2000) & Dec(J2000) & m$_{nIR}$ &M$_{nIR}$ &  r$_{80}$ (nIR)\\ 
		& & & &    AB mag  &    AB mag   & kpc \\
 		\hline
		\hline
		SN2006aj 	 & 0.03 & 	03:21:39.670 & +16:52:02.27  & 19.702 $\pm$  0.002 		&-16.250  $\pm$ 0.002 & 1.1 $\pm$ 0.4 \\
		SN2002hs 	 & 0.39 & 	03:32:18.590 & - 27:48:33.70 & 22.362 $\pm$  0.009 		&-18.897  $\pm$ 0.009 &1.1 $\pm$ 0.3\\
		SN2002fv 	 & 0.70 & 	03:32:19.220 & - 27:49:34.00 & 23.971 $\pm$  0.016 		&-18.608  $\pm$ 0.016 & 0.56 $\pm$ 0.13 \\
		SN2002hq 	 & 0.67 & 	03:32:29.940 & - 27:43:47.20 & 19.162 $\pm$  0.001 		&-23.323  $\pm$ 0.001 & 1.4$\pm$ 0.5\\	
		SN2002kb  	 & 0.58 & 	03:32:42.441 & - 27:50:25.08 & 19.221 $\pm$  0.0014 	&-22.9378 $\pm$ 0.0014 & 1.4 $\pm$ 0.5\\
		SN2002fz 	 & 0.84 & 	03:32:48.598 & - 27:54:17.14 & 20.385 $\pm$  0.002  	&-22.600  $\pm$0.002  & 0.9 $\pm$0.3\\
		SN2003ba 	 & 0.29 & 	12:36:15.925 &	+62:12:37.38 & 18.8852$\pm$  0.0009 	&-21.7071 $\pm$0.0009  & 1.2 $\pm$ 0.4\\
		SN2003bb 	 & 0.96 & 	12:36:24.506 &	+62:08:34.84 & 19.3208$\pm$  0.0016 	&-23.9609 $\pm$ 0.0016 & 1.3 $\pm$ 0.4\\
		SN2003ew 	 & 0.58 & 	12:36:27.828 &	+62:11:24.71 & 20.817 $\pm$  0.005 		& -21.346 $\pm$ 0.005&  1.3 $\pm$ 0.4\\
		SN2003dx 	 & 0.51 &  	12:36:31.772 &	+62:08:48.25 & 22.223 $\pm$  0.004 		&-19.648  $\pm$ 0.004 &  0.56 $\pm$ 0.15\\
		SN2003er 	 & 0.63 &  	12:36:32.270 &	+62:07:35.20 & 19.1932$\pm$  0.0006 	&-23.1551 $\pm$ 0.0006 &  0.9 $\pm$ 0.3\\
		SN2003en 	 & 0.54 &	12:36:33.179 &	+62:13:47.34 & 21.91  $\pm$  0.11 		&-20.09   $\pm$ 0.11 & 0.60 $\pm$ 0.16\\
		SN2003bc 	 & 0.51 &	12:36:38.130 &	+62:09:52.88 & 20.807 $\pm$  0.0018 	&-21.0661 $\pm$ 0.0018 & 0.8 $\pm$ 0.3\\
		SN2003dz 	 & 0.48 &	12:36:39.967 &	+62:07:52.12 & 23.81  $\pm$  0.03 		&-17.93  $\pm$ 0.03 &  0.63 $\pm$ 0.16\\
		SN2003N 	 & 0.43 &	12:37:09.140 &	+62:11:01.20 & 22.809 $\pm$  0.008 		&-18.677 $\pm$ 0.008 &  0.66 $\pm$ 0.18\\
		SN2003ea 	 & 0.98 &	12:37:12.066 &	+62:12:38.04 & 22.870 $\pm$  0.009 		&-20.457 $\pm$ 0.009 &  0.71 $\pm$0.18\\
		SN2002kl 	 & 0.41 &	12:37:49.350 &	+62:14:05.71 & 22.2   $\pm$  0.2 		&-19.20  $\pm$ 0.2 & 0.57 $\pm$ 0.15\\
		\hline
 		\end{tabular}
	\end{table*}
	
	\begin{table*}
	\centering
	\caption{ LGRB subsample from SNAPSHOT survey}
	\label{table:IR_LGRB_Comp}
	\begin{tabular}{llll}
	\hline
	Event & Redshift & RA(J2000) & Dec(J2000) \\
	\hline
	\hline
		GRB050824	&	0.828	& 00:48:56.260  & +22:36:33.20  \\ 
		GRB051016B	&	0.9364	& 08:48:27.860	& +13:39:19.60 \\  
		GRB060218	&	0.0331	& 03:21:39.650  & +16:52:01.30 \\  
		GRB060505	&	0.089	& 22:07:03.380	& -27:48:52.90 \\  
		GRB060602A	&	0.787	& 16:03:42.500	& +66:36:02.60  \\ 
		GRB060614	&	0.125	& 21:23:32.190	& -53:01:36.50 \\  
		GRB060729	&	0.54	& 06:21:31.840	& -62:22:12.10  \\ 
		GRB060912A	&	0.937	& 00:21:08.110	& +20:58:19.20   \\
		GRB061007	& 	1.2622	& 03:05:19.59	& -50:30:02.3	\\
		GRB061110A	&	0.758	& 22:25:09.850	& -02:15:31.00  \\ 
		GRB070318	&	0.840	& 03:13:56.760	& -42:56:46.80  \\ 
		GRB070521	&	1.3500	& 16:10:38.62	& +30:15:22.1  \\
		GRB071010A	&	0.98	& 19:12:14.624	& -32:24:07.16  \\ 
		GRB071010B	&	0.947	& 10:02:09.240	& +45:43:49.70  \\ 
		GRB071112C	&	0.823	& 02:36:50.910	& +28:22:16.80   \\
		GRB071112	& 	1.1400	& 18:26:25.26 	& +47:04:30.00  \\
		GRB080430	&	0.767	& 11:01:14.660	& +51:41:07.80  \\ 
		GRB080520	& 	1.5457	& 18:40:46.37 	& -54:59:30.6  \\
		GRB080707	& 	1.2322	& 02:10:28.41 	& +33:06:34.5   \\
		GRB080805	& 	1.5042	& 20:56:53.47 	& -62:26:40.2  \\
		GRB080916A	&	0.689	& 22:25:06.360	& -57:01:22.90  \\ 
		GRB081007	&	0.5295	& 22:39:50.500	& -40:08:49.80  \\ 
		GRB090424	&	0.544	& 12:38:05.090	& +16:50:15.70   \\
		GRB090618	&	0.54	& 19:35:58.400	& +78:21:25.20  \\ 
		GRB091127	&	0.49	& 02:26:19.910	& -18:57:08.90 \\  
		GRB091208B	&	1.063	& 01:57:34.090	& +16:53:22.70  \\ 
		\hline
 	\end{tabular}
	\end{table*}

\begin{table*}
\centering
\caption{  Above: Core Collapse SNe drawn the $^{1}$ GOODS sample and from $^{2}$ \citet{Sanders2012} and $^{3}$ \citet{Lennarz2012}.
Below: LGRBs from Ghosts  $^{4}$ 	\citet{Savaglio2009} (and references therein), $^{5}$ \citet{Resmi2012},   $^{6}$\citet{Hjorth2012},$^{7}$\citet{Sollerman2007}, $^{8}$\citet{Perley2013},$^{9}$ \citet{Levan2007} , $^{10}$ \citet{Cool2007}, $^{11}$ \citet{Tanvir2010}, $^{12}$ \citet{Kruhler2011}, $^{13}$ \citet{McBreen2010},$^{14}$  \citet{Holland2010}, $^{15}$ \citet{Vergani2011}, $^{16}$ \citet{Starling2011}, $^{17}$ \citet{SDSSDR72009}, $^{18}$\citet{Perez-Ramirez2013}, $^{19}$  \citet{Elliott2013}, $^{20}$ \citet{Levan2014B}.}
\label{table:UV_Comp}
	\begin{tabular}{l l l l l l l}
	 	\hline
		Event & Redshift & RA(J2000) & Dec(J2000) &  m$_{UV}$ &  M$_{UV}$	& ref.\\
		 	  & 		  & 		       & 	             & AB mag 	  & AB mag   	& 	 \\
 		\hline
		\hline		
		SN2002fz    &	0.841	&    03:32:48.598 	 &   -27:54:17.14  & 	22.36    $\pm$   0.01  	&   -20.636  $\pm$  0.01   &   $^{3}$	\\
		SN2002hq    &	0.669	&    03:32:29.94     &	 -27:43:47.2   & 	22.455   $\pm$   0.021	&   -20.074  $\pm$  0.021  &   $^{1}$	\\
		SN2002if    &	0.321	&    01:50:04.51     &	 +00:00:26.4   & 	20.54    $\pm$   0.044 	&   -20.38   $\pm$  0.044  &   $^{3}$	\\ 
		SN2002kb    &	0.58	&    03:32:42.441 	 &   -27:50:25.08  & 	21.337   $\pm$   0.007	&   -20.839  $\pm$  0.007  &   $^{3}$	\\
		SN2002ke    &	0.577	&    03:31:58.77	 &	 -27:45:00.7   & 	22.883   $\pm$   0.019	&   -19.316  $\pm$  0.019  &   $^{1}$	\\
		SN2002kl    &	0.41	&    12:37:49.350 	 &   +62:14:05.71  & 	23.81    $\pm$   0.01  	&   -17.595  $\pm$  0.01   &   $^{3}$	\\
		SN2003ba    &	0.286	&    12:36:15.925 	 &   +62:12:37.38  & 	21.533   $\pm$   0.197	&   -19.062  $\pm$  0.197  &   $^{3}$	\\
		SN2003bb    &	0.954	&    12:36:24.506 	 &   +62:08:34.84  & 	21.444   $\pm$   0.007	&   -21.836  $\pm$  0.007  &   $^{3}$	\\
		SN2003bc    &	0.511	&    12:36:38.130    & 	 +62:09:52.88  & 	22.645   $\pm$   0.008	&   -19.281  $\pm$  0.008  &   $^{1}$	\\
		SN2003dx    &	0.46	&    12:36:31.772 	 &   +62:08:48.25  & 	23.917   $\pm$   0.343	&   -17.745  $\pm$  0.343  &   $^{3}$	\\
		SN2003ea    &	0.89	&    12:37:12.066 	 &   +62:12:38.04  & 	24.13    $\pm$   nan   	&   -19.016  $\pm$  nan    &   $^{3}$	\\
		SN2003ew    &	0.66	&    12:36:27.828 	 &   +62:11:24.71  & 	22.603   $\pm$   0.193	&   -19.874  $\pm$  0.193  &   $^{3}$	\\
		HST04Geo    &	0.937	&    12:36:44.432    & 	 +62:10:53.19  & 	24.438   $\pm$   0.03 	&   -18.842  $\pm$  0.03   &   $^{1}$	\\
		HST04Riv    &	0.606	&    03:32:32.407    & 	 -27:44:52.84  & 	26.992   $\pm$   0.175	&   -15.315  $\pm$  0.175  &   $^{1}$	\\
		HST05Bra    &	0.48	&    12:37:21.764    & 	 +62:12:25.67  & 	23.649   $\pm$   0.023	&   -18.156  $\pm$  0.023  &   $^{1}$	\\
		HST05Den    &	0.971	&    12:37:14.773    & 	 +62:10:32.61  & 	25.949   $\pm$   0.106	&   -17.408  $\pm$  0.106  &   $^{1}$	\\
		SN2005hm    &	0.035	&    21:39:00.65     &	 -01:01:38.7   & 	21.5    $\pm$   0.22   &   -14.599  $\pm$  0.22   &   $^{2}$	\\
		SN2005nb    &	0.023	&    12:13:37.61     &	 +16:07:16.2   & 	15.966   $\pm$   0.011	&   -19.215  $\pm$  0.011  &   $^{2}$	\\
		SN2006ip    &	0.030	&    23:48:31.68     &	 -02:08:57.3   & 	17.263   $\pm$   0.022	&   -18.459  $\pm$  0.022  &   $^{2}$	\\
		SN2006ir    &	0.02	&    23:04:35.68     &	 +07:36:21.5   & 	17.347   $\pm$   0.027	&   -17.491  $\pm$  0.027  &   $^{2}$	\\
		SN2006jo    &	0.076	&    01:23:14.72     &	 -00:19:46.7   & 	18.073   $\pm$   0.028	&   -19.676  $\pm$  0.028  &   $^{2}$	\\
		SN2006nx    &	0.137	&    03:33:30.63     &	 -00:40:38.2   & 	21.119   $\pm$   0.192	&   -18.285  $\pm$  0.192  &   $^{2}$	\\
		SN2006sg    &	0.44	&    02:08:13.041 	 &   -03:46:21.93  & 	22.991   $\pm$   0.259	&   -18.608  $\pm$  0.259  &   $^{2}$	\\
		SN2006tq    &	0.26	&    02:10:00.698 	 &   -04:06:00.91  & 	22.855   $\pm$   0.617	&   -17.576  $\pm$  0.617  &   $^{2}$	\\
		SN2007I	    &	0.021	&    11:59:13.15     &	 -01:36:18.9   & 	19.11    $\pm$   0.07  	&   -15.827  $\pm$  0.07   &   $^{2}$	\\
		SN2007ea    &	0.04	&    15:53:46.27     &	 -27:02:15.5   & 	15.49    $\pm$   nan   	&   -20.715  $\pm$  nan    &   $^{2}$	\\
		SN2007ff    &	0.05	&    01:24:10.24     &	 +09:00:40.5   & 	17.322   $\pm$   0.027	&   -19.563  $\pm$  0.027  &   $^{2}$	\\
		SN2007gl    &	0.03	&    03:11:33.21     &	 -00:44:46.7   & 	17.057   $\pm$   0.026	&   -19.183  $\pm$  0.026  &   $^{2}$	\\
		SN2007hb    &	0.02	&    02:08:34.02     &	 +29:14:14.3   & 	15.617   $\pm$   0.009	&   -19.283  $\pm$  0.009  &   $^{2}$	\\
		SN2007hn    &	0.03	&    21:02:46.85     &	 -04:05:25.2   & 	18.295   $\pm$   0.036	&   -17.577  $\pm$  0.036  &   $^{2}$	\\
		SN2010ah    &	0.049	&    11:44:02.99     &	 +55:41:27.6   & 	20.15    $\pm$   0.12  	&   -16.544  $\pm$  0.12   &   $^{2}$	\\	
		\hline	
		GRB970228	&	0.695  &	 05:01:46.7   &		 +11:46:53 		& 25.1  $\pm$ 0.23 	& -18.2  $\pm$ 0.2 &   $^{4}$\\	 
		GRB970508 	&	0.8350 &	 06:53:49.2   &		 +79:16:19 		& 25.59 $\pm$0.15 	& -17.56  $\pm$ 0.15 & $^{4}$\\
		GRB970828 	&	0.9580 &	 18:08:31.6   &		 +59:18:51 		& 25.28 $\pm$ 0.29 	& -18.1  $\pm$ 0.3 &   $^{4}$\\
		GRB980425 	&	0.0085 &	 19:35:03.2   &		 -52:50:46		& 15.77 $\pm$ 0.03 	& -17.46  $\pm$ 0.03 & $^{4}$\\
		GRB980703 	&	0.9660 &	 23:59:06.7   &		 +08:35:07  	& 22.57 $\pm$0.06 	& -20.86  $\pm$ 0.06 & $^{4}$\\
		GRB990705 	&	0.842  &	 05:09:54.5   &		 -72:07:53		& 22.79 $\pm$ 0.18 	& -20.42  $\pm$ 0.18 & $^{4}$\\ 
		GRB990712 	&	0.434  &	 22:31:53.061 &		 -73:24:28.58 	& 23.15 $\pm$0.08 	& -18.46  $\pm$ 0.08 & $^{4}$\\
		GRB991208 	&	0.706  &	 16:33:53.51  &		 +46:27:21.5    & 24.51 $\pm$ 0.15 	& -18.15 $\pm$ 0.15 &  $^{4}$\\
		GRB000210 	&	0.846  &	 01:59:15.6   &		 -40:39:33 		& 24.18 $\pm$0.08 	& -18.89  $\pm$ 0.08 & $^{4}$\\
		GRB010921 	&	0.435  &	 22:55:59.90  &		 +40:55:52.9 	& 22.6  $\pm$ 0.1 	& -19.4  $\pm$ 0.1 &   $^{4}$\\
		GRB011121 	&	0.360  &	 11:34:26.67  & 	-76:01:41.6 	& 24.1  $\pm$0.1 	& -18.7  $\pm$ 0.1 &   $^{4}$\\ 
		GRB020405 	&	0.698  &	 13:58:03.12  &	 	 -31:22:22.2 	& 22.6  $\pm$0.05 	& -20.14  $\pm$ 0.05 & $^{4}$\\ 
		GRB020819B 	&	0.41   &	 23:27:19.475 & 	 +06:15:55.95 	& 20.31 $\pm$0.02 	& -21.33  $\pm$ 0.02 & $^{4}$\\
		GRB020903 	&	0.25   &	 22:48:42.34  &	 	 -20:46:09.3 	& 21.6  $\pm$ 0.09 	& -18.79  $\pm$ 0.09 & $^{4}$\\
		GRB030329 	&	0.168  &	 10:44:50.030 & 	 +21:31:18.15 	& 23.33 $\pm$ 0.09 	& -16.16  $\pm$ 0.09 & $^{4}$\\ 
		GRB030528 	&	0.782  &	 17:04:00.3	  & 	 -22:37:10	 	& 21.92 $\pm$ 0.18 	& -22.58  $\pm$ 0.18 & $^{4}$\\	 
		GRB031203 	&	0.1055 &	 08:02:30.4   &	 	 -39:51:00		& 18.23 $\pm$ 0.17  & -24.70  $\pm$ 0.17 & $^{4}$\\	
		GRB040924 	&	0.859  &	 02:06:22.52  & 	 +16:08:48.8 	& 24.31 $\pm$ 0.28  & -18.9  $\pm$ 0.3 &   $^{4}$\\ 
		GRB050525 	&	0.606  &	 18:32:32.560 & 	 +26:20:22.34 	& $\geq$24.0 		& $\geq$-18.586 &      $^{5}$\\
		GRB050824 	&	0.8278 &	 00:48:56.100 &	     +22:36:32.00  	& 23.77  $\pm$ 0.14 & -19.28 $\pm$ 0.14 &  $^{6,7}$\\
		\hline 
	 	\end{tabular}

\end{table*}

\begin{table*}
\centering
	\begin{tabular}{l l l l l l l}
	 	\hline
		Event & Redshift & RA(J2000) & Dec(J2000) &  m$_{UV}$&  M$_{UV}$ & ref.\\
		 	 & 		  & 		       & 		     & AB mag	& AB mag     &  \\ 
 		\hline
		\hline
		GRB050826 	&	0.296  &	 05:51:01.590  &	 	-02:38:35.40 & 21.37 $\pm$ 0.28 & -21.34 $\pm$ 0.28 &  $^{4}$\\
		GRB060202 	&	0.785  &	 02:23:22.940  &	 	+38:23:03.70 & 23.29 $\pm$ 0.07 & -19.72 $\pm$ 0.07 &  $^{8}$\\
		GRB060218 	&	0.0335 &	 03:21:39.670  & 	 	+16:52:0 	 & 20.5  $\pm$ 0.13 & -15.85 $\pm$ 0.13 & $^{4}$\\ 
		GRB060912A 	&	0.937  &	 00:21:08.11   &	 	+20:58:18.9  & 22.72 $\pm$ 0.04 & -20.63 $\pm$ 0.04 & $^{9}$\\
		GRB070612 	&	0.6710 &	 08:05.4	   & 		+37:15		 & 22.48 $\pm$ 0.17 & -20.19 $\pm$ 0.17 &  $^{10}$\\ 
		GRB080319B 	&	0.93   &	 14:31:41.04   &	 	+36:18:09.2  & 26.95 $\pm$ 0.12 & -16.29 $\pm$ 0.12 &  $^{11}$\\
		GRB081109	&	0.979  &	 22:03:11.50   &	 	-54:42:40.5  & 22.69 $\pm$ 0.06 & -20.69 $\pm$ 0.06 & $^{12}$\\
		GRB090328 	&	0.7354 &	 06:02:39.69   &		-41:52:55.1  & 22.64 $\pm$ 0.13 & -20.26 $\pm$ 0.13 &  $^{13}$\\
		GRB090417B 	&	0.345  &	 13:58:44.8	   &	 	+47:00:55 	 & 23.24 $\pm$ 0.53 & -17.8  $\pm$ 0.5  & $^{14}$\\
		GRB091127 	&	0.49   &	 02:26:19.87   &		-18:57:08.6  & 24.14 $\pm$ 0.16 & -17.79 $\pm$ 0.16 &  $^{15}$\\
		GRB100316D	&	0.0591 &	 07:10:30.63   &		-56:15:19.7  & 18.73 $\pm$ 0.09 & -18.97 $\pm$ 0.09 &  $^{16}$\\ 
		GRB100418 	&	0.6239 &	 17:05:26.96   &		+11:27:41.9  & 22.61 $\pm$ 0.16 & -19.97 $\pm$ 0.16 &  $^{17}$\\
		GRB100621A 	&	0.5420 &	 21:01:13.12   &		-51:06:22.5  & 21.79 $\pm$ 0.06 & -20.34 $\pm$ 0.06 & $^{12}$\\
		GRB100816A 	&	0.8049 &	 23:26:57.56   &	 	+26:34:42.6  & 23.08 $\pm$ 0.15 & -20.02 $\pm$ 0.15 &  $^{18}$\\
		GRB110918 	&	0.984  &	 02:10:09.39   &	 	-27:06:19.6  & 22.04 $\pm$ 0.05 & -21.35 $\pm$ 0.05 &  $^{19}$\\
		GRB101225A 	&	0.85   &	 00:00:47.48   &		+44:36:01.0  & 26.75 $\pm$ 0.13 & -16.60 $\pm$ 0.13 &  $^{20}$\\
		GRB111209A 	&	0.67   &	 00:57:22.700  &	    -46:48:05.00 & 25.75 $\pm$ 0.14 & -16.82 $\pm$ 0.14 & $^{20}$\\
		GRB120422A 	&	0.28   &	 09:07:38.38   &		+14:01:07.5  & 22.17 $\pm$ 0.5  & -18.5  $\pm$ 0.5  & $^{17}$\\
		GRB130427A 	&	0.35   &	 11:32:32.63   &		+27:41:51.7  & 22.84 $\pm$ 0.08 & -18.29 $\pm$ 0.08 & $^{20}$\\
		\hline
 	\end{tabular}
\end{table*}

\bsp	
\label{lastpage}
\end{document}